\documentclass[a4paper,11pt]{article}
\baselineskip
\usepackage[T1]{fontenc}
\usepackage{amsmath}
\usepackage{mathtools}
\usepackage{subcaption}
\usepackage{amssymb}
\usepackage{graphicx}
\usepackage{latexsym}
\usepackage{cite}
\usepackage{collref}
\usepackage{color}
\usepackage{hyperref}
\DeclareMathOperator{\arctanh}{arctanh}
\DeclareMathOperator{\Tr}{Tr}

\newcommand{\Z}{\mathbb{Z}}
\newcommand{\SU}{\mathrm{SU}}

\newcommand{\dd}{{\rm{d}}}
\newcommand{\Zp}{Z_{\mbox{\tiny{p}}}}
\newcommand{\Za}{Z_{\mbox{\tiny{a}}}}
\newcommand{\Fin}{F_{\mbox{\tiny{in}}}}
\newcommand{\Ffin}{F_{\mbox{\tiny{fin}}}}
\newcommand{\tin}{t_{\mbox{\tiny{in}}}}
\newcommand{\tfin}{t_{\mbox{\tiny{fin}}}}
\newcommand{\Tin}{T_{\mbox{\tiny{in}}}}
\newcommand{\Tfin}{T_{\mbox{\tiny{fin}}}}
\newcommand{\Tc}{T_{\mbox{\tiny{c}}}}
\newcommand{\kB}{k_{\mbox{\tiny{B}}}}
\newcommand{\betagauge}{\beta_{\mbox{\tiny{g}}}}
\newcommand{\effaction}{\Gamma_{\mbox{\tiny{eff}}}}
\newcommand{\Scl}{S_{\mbox{\tiny{cl}}}}
\newcommand{\nr}{n_{\mbox{\tiny{r}}}}
\newcommand{\Seff}{S_{\mbox{\tiny{eff}}}}
\newcommand{\redchisq}{\chi^2_{\tiny\mbox{red}}}
\begin{document}

\begin{titlepage}
\vskip0.5cm 
\begin{flushright} 
CP3-Origins-2016-020 DNRF90\\DIAS-2016-20
\end{flushright} 
\vskip0.5cm 
\begin{center}
{\Large\bf Jarzynski's theorem for lattice gauge theory}
\end{center}
\vskip1.3cm
\centerline{Michele~Caselle$^{a}$, Gianluca~Costagliola$^{a}$, Alessandro~Nada$^{a}$,}
\centerline{Marco~Panero$^{a}$, and Arianna~Toniato$^{a,b}$}
\vskip1.5cm
\centerline{\sl $^a$ Department of Physics, University of Turin \& INFN, Turin}
\centerline{\sl Via Pietro Giuria 1, I-10125 Turin, Italy}
\vskip0.5cm
\centerline{\sl $^b$ CP$^3$-Origins \& Danish IAS, University of Southern Denmark}
\centerline{\sl Campusvej 55, 5230 Odense M., Denmark}
\vskip0.5cm
\begin{center}
{\sl  E-mail:} \hskip 1mm \href{mailto:caselle@to.infn.it}{{\tt caselle@to.infn.it}}, \href{mailto:costagli@to.infn.it}{{\tt costagli@to.infn.it}}, \href{mailto:anada@to.infn.it}{{\tt anada@to.infn.it}}, \href{mailto:marco.panero@unito.it}{{\tt marco.panero@unito.it}} , \href{mailto:toniato@cp3.sdu.dk}{{\tt toniato@cp3.sdu.dk}}
\end{center}
\vskip1.0cm
\begin{abstract}
\noindent Jarzynski's theorem is a well-known equality in statistical mechanics, which relates fluctuations in the work performed during a non-equilibrium transformation of a system, to the free-energy difference between two equilibrium ensembles. In this article, we apply Jarzynski's theorem in lattice gauge theory, for two examples of challenging computational problems, namely the calculation of interface free energies and the determination of the equation of state. We conclude with a discussion of further applications of interest in QCD and in other strongly coupled gauge theories, in particular for the Schr\"odinger functional and for simulations at finite density using reweighting techniques.
\end{abstract}
\vspace*{0.2cm}

\end{titlepage}

\section{Introduction}
\label{sec:introduction}

Both in statistical mechanics and in quantum field theory, the numerical study of a large class of physical quantities by Monte~Carlo methods can be reduced to the evaluation of differences of free energies $F$. For lattice gauge theory, the most typical examples arise in the investigation of the phase diagram of QCD and QCD-like theories. For instance, in the study of the QCD equation of state at finite temperature $T$ (and zero baryon density), the difference between the pressure $p(T)$ and its value at $T=0$ can be computed using the fact that $p$ is opposite to the free energy density $f=F/V$, where $V$ denotes the system volume.\footnote{Strictly speaking, the $p=-f$ equality holds only for an infinite-volume system. In a periodic, cubic box of volume $V=L^3$, the relation is violated by corrections that depend on the aspect ratio $LT$ of the time-like cross-section of the hypertorus (for a gas of free, massless bosons)~\cite{Gliozzi:2007jh} or on the ratio of the linear size of the system $L$ over the inverse of the smallest screening mass (if screening effects are present)~\cite{DeTar:1985kx, Elze:1988zs, Meyer:2009kn}: see also ref.~\cite{Panero:2008mg} for a numerical study of these effects on lattices of typical sizes used in Monte~Carlo simulations.} In turn, $f$ can then be evaluated for example by ``integrating a derivative''~\cite{Engels:1990vr}: during the past few years, this method has led to high-precision determinations of the equation of state for QCD~\cite{Bazavov:2009zn, Borsanyi:2013bia} and for Yang--Mills theories based on different gauge groups~\cite{Umeda:2008bd, Panero:2009tv, Borsanyi:2012ve, Bruno:2014rxa} and/or in lower dimensions~\cite{Bialas:2008rk, Caselle:2011mn}. These results can be compared with those obtained in other recent works~\cite{Asakawa:2013laa, Giusti:2014ila}, in which novel techniques (respectively based on the Wilson flow~\cite{Luscher:2010iy, Suzuki:2013gza} and on shifted boundary conditions~\cite{Giusti:2010bb, Giusti:2012yj}) have been used.

Other objects having a natural interpretation in terms of free-energy differences in finite-temperature non-Abelian gauge theories are the interfaces separating different center domains and/or regions of space characterized by different realizations of center symmetry~\cite{Kajantie:1988hn, Kajantie:1989xk, Enqvist:1990ae, Huang:1990jf, Kajantie:1990bu, Bhattacharya:1990hk, Bhattacharya:1992qb, KorthalsAltes:1993ca, Iwasaki:1993qu, Monden:1997hb, Giovannangeli:2001bh, Pisarski:2002ji}: they could have phenomenological implications for heavy-ion collisions~\cite{Asakawa:2012yv} and for cosmology~\cite{Ignatius:1991nk} and have been studied quite extensively in lattice simulations~\cite{Lucini:2003zr, Lucini:2005vg, Bursa:2005yv}.

In the study of QCD at finite baryon chemical potential $\mu$, a possible computational strategy to cope with the notorious sign problem~\cite{deForcrand:2010ys, Philipsen:2012nu, Levkova:2012jd, Aarts:2013lcm, D'Elia:2015rwa, Gattringer:2016kco} is the one based on the method first introduced in ref.~\cite{Ferrenberg:1988yz} and later extended to applications in lattice QCD~\cite{Toussaint:1989fn, Fodor:2001au}, whereby importance sampling is carried out in an ensemble of configurations generated using the determinant of the Dirac operator $D$ at $\mu=0$, and the expectation value in the target ensemble at finite $\mu$ is obtained through reweighting by the expectation value of $\det D(\mu)/\det D(0)$, computed in the $\mu=0$ ensemble. The natural logarithm of the latter quantity can be interpreted as ($1/T$ times) the difference between the free energies associated with the partition functions of the $\mu=0$ and finite-$\mu$ ensembles. Note that the \emph{extensive} nature of these quantities implies that a severe overlap problem arises in a large volume: for a Markov chain generated using the determinant of the Dirac operator $D$ at $\mu=0$, the probability of probing those regions of phase space, where the measure of the finite-$\mu$ ensemble is largest, gets exponentially suppressed with the system hypervolume, resulting in extremely poor sampling.

Free-energy differences are also relevant for the study of operators in the ground state of gauge theories. For example, vacuum expectation values of extended operators like 't~Hooft loops ($\widetilde{\mathcal{W}}$)~\cite{'tHooft:1977hy}, which have been studied on the lattice in several works~\cite{Kovacs:2000sy, Hoelbling:2000su, DelDebbio:2000cx, deForcrand:2000fi, deForcrand:2001nd}, can be generically written in the form
\begin{equation}
\label{tHooft}
\langle \widetilde{\mathcal{W}} \rangle = \frac{\int \mathcal{D} \phi \widetilde{\mathcal{W}}[\phi] \exp \left(-S[\phi]\right)}{\int \mathcal{D} \phi \exp \left(-S[\phi]\right)} = \frac{Z_{\widetilde{\mathcal{W}}}}{Z} = \exp\left[ - \left( F_{\widetilde{\mathcal{W}}} - F \right) L \right],
\end{equation}
where $\mathcal{D} \phi$ denotes the measure for the (regularized) functional integration over the generic fields $\phi$, $S$ is the Euclidean action, $Z$ is the partition function, $F$ is the free energy, and $L$ is the system size in the Euclidean-time direction, while $Z_{\widetilde{\mathcal{W}}}$ denotes a \emph{modified} partition function, in which the observable has been included in the action (by twisting a set of plaquettes that tile the $\widetilde{\mathcal{W}}$ loop~\cite{Srednicki:1980gb}) and $F_{\widetilde{\mathcal{W}}}$ is the corresponding free energy. Note that, in the case of a ``maximal'' 't~Hooft loop, i.e. one extending through a whole cross-section of the system, this problem has a natural connection to the study of fluctuating interfaces in statistical mechanics. It is worth noting that there exist many experimental realizations of fluctuating interfaces, particularly in mesoscopic physics, in chemistry and in biophysics: some well-known examples include binary mixtures and amphiphilic membranes~\cite{Gelfand:1990fse, Privman:1992zv}.

Other extended operators, like Wilson loops or Polyakov-loop correlation functions, can be easily recast into simple expressions of the form of eq.~(\ref{tHooft}) in a dual formulation of the theory, at least for Abelian (or, more generally, solvable) gauge groups~\cite{Panero:2004zq, Panero:2005iu, Caselle:2014eka, Caselle:2016mqu}.

This list of examples is by no means exhaustive, as the class of physical observables whose expectation values can be written in a natural way in terms of a free-energy difference---i.e. as a ratio of partition functions---is much broader. Note that, while it is always possible to trivially \emph{define} the expectation value of any arbitrary operator $\mathcal{O}$ as a ratio of partition functions of the form $Z_{\mathcal{O}}/Z=\exp\left[-\left(F_{\mathcal{O}}-F\right)L\right]$, here we are interested in the cases in which the quantity $Z_{\mathcal{O}}$ can be written as an integral over positive weights, that can be sampled efficiently by Monte~Carlo methods.

The examples above (and the computational problems that they involve) show that, in general, the numerical evaluation of free-energy differences remains a non-trivial computational challenge---one that cannot be easily tackled by brute-force approaches---in particular for large systems.

In this work, we present an application of non-equilibrium methods from numerical statistical mechanics, in lattice gauge theory. More precisely, we show that the class of algorithms based on Jarzynski's relation (whose derivation is presented in section~\ref{sec:Jarzynski}, along with some comments relevant for practical implementations in Monte~Carlo simulations) can be applied to gauge theories formulated on a Euclidean lattice, in a straightforward way. In a nutshell, this is so, because the Euclidean lattice formulation of a gauge theory~\cite{Wilson:1974sk} can be interpreted as a statistical mechanics system of a countable (and, in actual Monte~Carlo simulations, finite) number of degrees of freedom~\cite{Kogut:1979wt}. The main difference of Euclidean lattice gauge theories with respect to statistical spin models, namely the existence of an invariance under \emph{local}, rather than \emph{global}, transformations of the internal degrees of freedom, does not play any r\^ole in Jarzynski's theorem, so that there is no conceptual obstruction to its application for lattice gauge theories. Nevertheless, this theorem has received surprisingly little attention in the lattice community. With the notable exception of some works carried out in the three-dimensional Ising model (see, e.g., ref.~\cite{Chatelain:2007ts} and additional references mentioned below), which is exactly equivalent to a three-dimensional $\Z_2$ lattice gauge theory, we are not aware of any large-scale numerical studies of lattice QCD or of other lattice gauge theories, using Jarzynski's theorem. A motivation of the present work is to partially fill this gap, by presenting examples of applications of Jarzynski's theorem in two computationally challenging problems, and, as will be discussed in more detail in the following, by initiating a study of the practical details of computationally efficient algorithmic implementations of Jarzynski's relation. We will discuss applications in two different problems, namely in a high-precision numerical study of the physics of fluctuating interfaces, and in the calculation of the equation of state in non-Abelian gauge theories. The body of literature about the dynamics of interfaces (in different statistical-mechanics models) is vast~\cite{Binder:1982mc, Burkner:1983mc, Berg:1991sn, Hasenbusch:1992zz, Potoff:2000st, Davidchack:2005cs, Caselle:1992ue, Caselle:1994df, Caselle:2006dv, Billo:2006zg, Caselle:2007yc, Billo:2007fm,
condmat0602580, Chatelain:2007ts, Hijar2007, 0905.4569, Limmer:2011tp, Binder:2011mc, 1401.7870, 1406.0616, 1411.5588}; for our present purposes, particularly relevant works include those that have been recently carried out by Binder and collaborators (see refs.~\cite{1401.7870, 1406.0616, 1411.5588} and references therein), as well as those reported in refs.~\cite{condmat0602580, Chatelain:2007ts, Hijar2007}. We will also compare our new results with those obtained in earlier works by the Turin group~\cite{Caselle:1992ue, Caselle:1994df, Caselle:2006dv, Billo:2006zg, Caselle:2007yc, Billo:2007fm}. The results obtained in this benchmark study are compared with state-of-the-art analytical predictions based on an effective-string model~\cite{Aharony:2009gg, Aharony:2010cx, Aharony:2010db, Kol:2010fq, Aharony:2011gb, Gliozzi:2012cx, Dubovsky:2012sh, Aharony:2013ipa, Caselle:2013dra, Ambjorn:2014rwa, Brandt:2016xsp}: the precision of the results that we obtain with this algorithm in $\Z_2$ gauge theory in three dimensions allows us to clearly resolve subleading corrections predicted by the effective theory, which scale like the \emph{seventh} and the \emph{ninth} inverse powers of the linear size of the interface. In section~\ref{sec:equation_of_state} we discuss an implementation of this type of algorithm in non-Abelian gauge theory with $\SU(2)$ gauge group, and present preliminary results for the computation of the equation of state in the confining phase of this theory. Finally, in section~\ref{sec:conclusions} we summarize the key features of non-equilibrium algorithms like the one discussed in this work, and discuss their potential for applications in computationally challenging problems, in particular those relevant for the calculation of free energies (or, more generally, effective actions) in QCD and in other strongly coupled field theories.

\section{Jarzynski's relation}
\label{sec:Jarzynski}

The class of algorithms that we are discussing in the present work are based on a theorem proven by Jarzynski in refs.~\cite{Jarzynski:1996ne, Jarzynski:1997ef} (for a discussion about the relation with earlier work by Bochkov and Kuzovlev~\cite{Bochkov:1977gt, Bochkov:1979fd, Bochkov:1981nf}, see refs.~\cite{condmat0612305, Kuzovlev:2011sr}; for the connection with entropy-production fluctuation theorems~\cite{Evans:1993po}, see ref.~\cite{Crooks:1999ep}). Remarkably, this relation has also been verified experimentally, as discussed, for instance, in ref.~\cite{Liphardt:2002ei}.

In a nutshell, Jarzynski's relation states the equality of the exponential average of the work done on a system in non-equilibrium processes, and the difference between the free energies of the initial ($\Fin$) and the final ($\Ffin$) ensembles, respectively associated with the system parameters realized at ``times'' $\tin$ and $\tfin$. Here, ``time'' can either refer to Monte~Carlo time (in a numerical simulation), or to real time (in an experiment), and the average is taken over a large number of realizations of such non-equilibrium evolutions from the initial and the final ensembles.

In the following, we summarize the original derivation presented in refs.~\cite{Jarzynski:1996ne, Jarzynski:1997ef}, using natural units ($\hbar=c=\kB=1$) and focusing, for definiteness, on a statistical-mechanics system---although, as we will show below, the generalization to lattice gauge theories is straightforward.

Consider a system, whose microscopic degrees of freedom are collectively denoted as $\phi$ (for instance, $\phi$ could represent the spins defined on the sites of a regular $D$-dimensional lattice: $\phi = \{ \phi_{(x_1,\dots,x_D)}\}$). Let the dynamics of the system be described by the Hamiltonian $H$, which is a function of the degrees of freedom $\phi$, and depends on a set of parameters (e.g. couplings). When the system is in thermal equilibrium with a large heat reservoir at temperature $T$, the partition function of the system is
\begin{equation}
\label{partition_function}
Z =\sum_{\phi} \exp \left( - \frac{H}{T} \right),
\end{equation}
where, as usual, $\sum_{\phi}$ denotes the multiple sum (or integral) over the values that each microscopic degree of freedom can take. The statistical distribution of $\phi$ configurations in thermodynamic equilibrium is given by the Boltzmann distribution:
\begin{equation}
\label{Boltzmann_distribution}
\pi[\phi] = \frac{1}{Z} \exp \left( - \frac{H}{T} \right),
\end{equation}
which, in view of eq.~(\ref{partition_function}), is normalized to $1$:
\begin{equation}
\label{Boltzmann_distribution_normalization}
\sum_\phi \pi[\phi] = 1.
\end{equation}
Let us denote the conditional probability (or the conditional probability density, if the degrees of freedom of the system can take values in a continuous domain) that the system undergoes a transition from a configuration $\phi$ to a configuration $\phi^\prime$ as $P[\phi\to\phi^\prime]$. The sum of such probabilities over all possible distinct final configurations is one,
\begin{equation}
\label{transition_probability}
\sum_{\phi^\prime} P[\phi\to\phi^\prime] = 1,
\end{equation}
because the system certainly must evolve to \emph{some} final configuration. Since the Boltzmann distribution is an equilibrium thermal distribution, it satisfies the property
\begin{equation}
\label{Boltzmann_distribution_stationarity}
\sum_{\phi} \pi[\phi] P[\phi\to\phi^\prime] = \pi[\phi^\prime].
\end{equation}
In the following, we will assume that the system satisfies the stronger, detailed-balance condition:
\begin{equation}
\label{detailed_balance}
\pi[\phi] P[\phi\to\phi^\prime] = \pi[\phi^\prime] P[\phi^\prime \to \phi];
\end{equation}
note that, if eq.~(\ref{transition_probability}) holds, then eq.~(\ref{detailed_balance}) implies eq.~(\ref{Boltzmann_distribution_stationarity}), but the converse is not true.

In general, the Boltzmann distribution $\pi$ (as well as $Z$ and $P$) will depend on the couplings appearing on the Hamiltonian and on the temperature $T$; denoting them collectively as $\lambda$, one can then highlight such dependence by writing the configuration distribution as $\pi_\lambda$ (and the partition function and transition probabilities as $Z_\lambda$ and $P_\lambda$, respectively).

Let us introduce a time dependence for the $\lambda$ parameters---including the couplings of the Hamiltonian and, possibly, also the temperature $T$~\cite{Chatelain:2007ts}. Starting from a situation, at the initial time $t=\tin$, in which the couplings of the Hamiltonian take certain values, and the system is in thermal equilibrium at the temperature $\Tin$, the parameters of the system are modified as functions of time, according to some specified procedure, $\lambda(t)$, and are driven to final values $\lambda(\tfin)$ over an interval of time $\Delta t = \tfin-\tin$. $\lambda(t)$ is assumed to be a continuous function; for simplicity, we take it to interpolate linearly in $(t-\tin)$ between the initial, $\lambda(\tin)$, and final, $\lambda(\tfin)$, values. During the time interval between $\tin$ and $\tfin$, the system is, in general, out of thermal equilibrium.\footnote{For example, the parameters of the system could be changed in a sufficiently short interval of real time in an actual experiment, or of Monte~Carlo time in a numerical simulation. Unless $\Delta t \to \infty$, the system ``does not have enough time'' to thermalize.}

Now, discretize the $\Delta t$ interval in $N$ sub-intervals of the same width $\tau=\Delta t /N$, define $t_n=\tin + n \tau$ for integer values of $n$ ranging from $0$ to $N$ (so that $t_0=\tin$ and $t_N=\tfin$); correspondingly, the linear $\lambda(t)$ mentioned above can be discretized by a piecewise-constant function, taking the value $\lambda(t_n)$ for $t_n \le t < t_{n+1}$. Furthermore, let $\phi(t)$ denote one possible (arbitrary) ``trajectory'' in the space of field configurations, i.e. a mapping between the time interval $[\tin,\tfin]$ and the configuration space of the system; upon discretization of the $[\tin,\tfin]$ interval, such trajectory can be associated with the $(N+1)$-dimensional array of field configurations defined as $\left\{ \phi(\tin), \phi(t_1), \phi(t_2), \dots , \phi(t_{N-1}),  \phi(\tfin) \right\}$. Finally, let us introduce the quantity $\mathcal{R}_N[\phi]$ defined as
\begin{equation}
\label{discretized_exponential_work}
\mathcal{R}_N[\phi] = \exp \left( - \sum_{n=0}^{N-1} \left\{ \frac{H_{\lambda\left(t_{n+1}\right)}\left[\phi\left(t_n\right)\right]}{T\left(t_{n+1}\right)} - \frac{H_{\lambda\left(t_n \right)}\left[\phi\left(t_n \right)\right]}{T\left(t_n \right)} \right\}\right)
\end{equation}
(where the Hamiltonian $H_\lambda$ depends on its couplings, not on the temperature $T$): each summand appearing on the right-hand side of eq.~(\ref{discretized_exponential_work}) is the work (over $T$) done on the system during a time interval $\tau$, by switching the couplings from their values at $t=\tin + n \tau$ to those at $t=\tin + (n+1) \tau$. Thus, $\mathcal{R}_N[\phi]$ provides a discretization of the exponentiated work done on the system in the time interval from $t=\tin$ to $t=\tfin$, during which the parameters are switched as a function of time, $\lambda(t)$, and the fields trace out the trajectory $\phi(t)$ in configuration space. This discretization gets more and more accurate for larger and larger values of $N$, and becomes exact in the $N \to \infty$ limit, whereby the sum on the right-hand side of eq.~(\ref{discretized_exponential_work}) turns into a definite integral.

Recalling that the usual mapping between statistical mechanics and lattice gauge theory~\cite{Wilson:1974sk} associates $H/T$ with the Euclidean action of the lattice theory, one easily realizes that, from the point of view of the lattice theory, each term within the braces on the right-hand side of eq.~(\ref{discretized_exponential_work}) can be interpreted as the \emph{difference in Euclidean action} for the field configuration denoted as $\phi\left(t_n \right)$, which is induced when the parameters are changed from $\lambda\left(t_n \right)$ to $\lambda\left(t_{n+1}\right)$. Thus, evaluating the work (over $T$) during a Monte Carlo simulation of this statistical system corresponds to evaluating the variation in Euclidean action in the lattice gauge theory---and this is precisely the quantity that was evaluated in the simulations discussed in sections~\ref{sec:interface} and~\ref{sec:equation_of_state}.

Using eq.~(\ref{Boltzmann_distribution}), eq.~(\ref{discretized_exponential_work}) can then be recast in the form
\begin{equation}
\label{discretized_exponential_work_Z_pi_ratios}
\mathcal{R}_N[\phi] = \prod_{n=0}^{N-1} \frac{Z_{\lambda(t_{n+1})} \cdot \pi_{\lambda(t_{n+1})}\left[\phi\left(t_n \right)\right]}{Z_{\lambda(t_n)} \cdot \pi_{\lambda(t_n)}\left[\phi\left(t_n \right)\right]} .
\end{equation}
Next, consider the average of eq.~(\ref{discretized_exponential_work_Z_pi_ratios}) over all possible field-configuration trajectories realizing an evolution of the system from one of the configurations of the initial ensemble (at $t=\tin$, when the parameters of the system take the values $\lambda(\tin)$) to a configuration of the final ensemble (at $t=\tfin$, when the parameters of the system take the values $\lambda(\tfin)$). In practice, in a Monte Carlo simulation, this is realized by averaging over a sufficiently large number of discretized trajectories starting from configurations of the initial, equilibrium ensemble (described by the partition function $Z_{\lambda(\tin)}$ and by the canonical distribution $\pi_{\lambda(\tin)}$), and assuming that, given a configuration of fields at a certain time $t=t_n$, a new field configuration at time $t=t_{n+1}$ is obtained by Markov evolution with transition probability $P_{\lambda(t_{n+1})}\left[ \phi(t_n) \to \phi(t_{n+1}) \right]$, which is assumed to satisfy the detailed balance condition eq.~(\ref{detailed_balance}). Note that $P$ is taken to depend on $\lambda(t_{n+1})$: for every finite value of $\tau$ (and for every Monte Carlo computation with finite statistics), this way of discretizing the non-equilibrium transformation introduces an ``asymmetry'' in the time evolution (one could alternatively carry out the two steps in the opposite order) and a related systematic uncertainty. As it will be discussed below, this leads to a difference in the results obtained when the transformation of the parameters is carried out in one direction or in the opposite one, but such ``discretization effect'' is expected to vanish for $\tau \to 0$ (i.e. for $N \to \infty$), and our numerical results do confirm that. Another, more important, reason why the evolution of the system is not ``symmetric'' under time reversal, is that, while the initial ensemble is at equilibrium, this is not the case at later times: the system is progressively driven (more and more) out of equilibrium.

Then, the average of eq.~(\ref{discretized_exponential_work_Z_pi_ratios}) over all possible field-configuration trajectories realizing an evolution of the system from $t=\tin$ to  $t=\tfin$ can be written as
\begin{equation}
\label{averaged_discretized_exponential_work_Z_pi_ratios}
\sum_{\left\{ \phi(t) \right\} } \mathcal{R}_N[\phi] = \sum_{\left\{ \phi(t) \right\} } \pi_{\lambda(\tin)}\left[ \phi(\tin) \right] \prod_{n=0}^{N-1} \left\{ \frac{Z_{\lambda(t_{n+1})}}{Z_{\lambda(t_n)}}  \cdot \frac{\pi_{\lambda(t_{n+1})}\left[\phi\left(t_n \right)\right]}{\pi_{\lambda(t_n)}\left[\phi\left(t_n \right)\right]} \cdot  P_{\lambda(t_{n+1})}\left[ \phi(t_n) \to \phi(t_{n+1}) \right] \right\},
\end{equation}
where we used the fact that the system is initially in thermal equilibrium, hence the probability distribution for the configurations at $t=\tin$ is given by eq.~(\ref{Boltzmann_distribution}), and where $\sum_{\left\{ \phi(t) \right\} }$ denotes the $N+1$ sums over field configurations at all discretized times from $\tin$ to $\tfin$:
\begin{equation}
\sum_{\left\{ \phi(t) \right\} } \dotsc = \sum_{\phi(\tin)} \sum_{\phi(t_1)}  \sum_{\phi(t_2)} \dots \sum_{\phi\left(\tfin-\tau\right)} \sum_{\phi(\tfin)} \dotsc .
\end{equation}
The telescopic product of partition-function ratios in eq.~(\ref{averaged_discretized_exponential_work_Z_pi_ratios}) simplifies, and the equation can be rewritten as 
\begin{equation}
\label{discretized_exponential_work_pi_ratios}
\sum_{\left\{ \phi(t) \right\} } \mathcal{R}_N[\phi] = \frac{Z_{\lambda(\tfin)}}{Z_{\lambda(\tin)}} \sum_{\left\{ \phi(t) \right\} } \pi_{\lambda(\tin)}\left[ \phi(\tin) \right] \prod_{n=0}^{N-1} \left\{  \frac{\pi_{\lambda(t_{n+1})}\left[\phi\left(t_n \right)\right] \cdot  P_{\lambda(t_{n+1})}\left[ \phi(t_n) \to \phi(t_{n+1}) \right]}{\pi_{\lambda(t_n)}\left[\phi\left(t_n \right)\right]}  \right\}.
\end{equation}
Using eq.~(\ref{detailed_balance}), this expression can be turned into
\begin{equation}
\label{simplified_discretized_exponential_work_pi_ratios}
\sum_{\left\{ \phi(t) \right\} } \mathcal{R}_N[\phi] = \frac{Z_{\lambda(\tfin)}}{Z_{\lambda(\tin)}} \sum_{\left\{ \phi(t) \right\} } \pi_{\lambda(\tin)}\left[ \phi(\tin) \right] \prod_{n=0}^{N-1} \left\{  \frac{\pi_{\lambda(t_{n+1})}\left[\phi\left(t_{n+1} \right)\right]}{\pi_{\lambda(t_n)}\left[\phi\left(t_n \right)\right]} \cdot  P_{\lambda(t_{n+1})}\left[ \phi(t_{n+1}) \to \phi(t_n)\right]  \right\}.
\end{equation}
At this point, also the telescopic product of ratios of Boltzmann distributions can be simplified, reducing the latter expression to
\begin{equation}
\label{discretized_exponential_P_product}
\sum_{\left\{ \phi(t) \right\} } \mathcal{R}_N[\phi] = \frac{Z_{\lambda(\tfin)}}{Z_{\lambda(\tin)}} \sum_{\left\{ \phi(t) \right\} } \pi_{\lambda(\tfin)}\left[ \phi(\tfin) \right] \prod_{n=0}^{N-1} P_{\lambda(t_{n+1})}\left[ \phi(t_{n+1}) \to \phi(t_n)\right].
\end{equation}
Note that, in eq.~(\ref{discretized_exponential_P_product}), $\phi(\tin)$ appears only in the $P_{\lambda(t_1)}\left[ \phi(t_1) \to \phi(\tin)\right]$ term: thus, one can use eq.~(\ref{transition_probability}) to carry out the sum over the $\phi(\tin)$ configurations, and eq.~(\ref{discretized_exponential_P_product}) reduces to
\begin{equation}
\label{simplified_discretized_exponential_P_product}
\sum_{\left\{ \phi(t) \right\} } \mathcal{R}_N[\phi] = \frac{Z_{\lambda(\tfin)}}{Z_{\lambda(\tin)}} \sum_{\phi(t_1)} \sum_{\phi(t_2)} \dots \sum_{\phi\left(\tfin-\tau\right)} \sum_{\phi(\tfin)} \pi_{\lambda(\tfin)}\left[ \phi(\tfin) \right] \prod_{n=1}^{N-1} P_{\lambda(t_{n+1})}\left[ \phi(t_{n+1}) \to \phi(t_n)\right].
\end{equation}
Repeating the same argument, eq.~(\ref{simplified_discretized_exponential_P_product}) can then be simplified using the fact that the only remaining dependence on $\phi(t_1)$ is in the $P_{\lambda(t_2)}\left[ \phi(t_2) \to \phi(t_1)\right]$ term, and so on. One arrives at
\begin{equation}
\label{almost_completely_simplified_discretized_exponential_P_product}
\sum_{\left\{ \phi(t) \right\} } \mathcal{R}_N[\phi] = \frac{Z_{\lambda(\tfin)}}{Z_{\lambda(\tin)}} \sum_{\phi(\tfin)} \pi_{\lambda(\tfin)}\left[ \phi(\tfin) \right].
\end{equation}
Finally, eq.~(\ref{Boltzmann_distribution_normalization}) implies that also the last sum yields one, so one gets
\begin{equation}
\label{almost_completely_simplified_discretized_exponential_P_product}
\sum_{\left\{ \phi(t) \right\} } \mathcal{R}_N[\phi] = \frac{Z_{\lambda(\tfin)}}{Z_{\lambda(\tin)}}.
\end{equation}
Recalling that, as we discussed above, in the large-$N$ limit $\mathcal{R}_N[\phi]$ equals the exponentiated work done on the system during the evolution from $\tin$ to $\tfin$, and writing $Z_{\lambda(\tin)}$ and $Z_{\lambda(\tfin)}$ in terms of the associated equilibrium free energies at the respective temperatures, eq.~(\ref{almost_completely_simplified_discretized_exponential_P_product}) yields the (generalized) Jarzynski relation:
\begin{equation}
\label{generalized_Jarzynski}
\left\langle \exp \left[ - \int \frac{\delta W}{T} \right] \right\rangle = \exp \left[ - \left( \frac{\Ffin}{\Tfin} - \frac{\Fin}{\Tin}\right)\right],
\end{equation}
where $\delta W$ denotes the work done on the system during an infinitesimal interval in the transformation from $\tin$ to $\tfin$, the integral is taken over all such intervals, and the average is taken over all possible realizations of this transformation.

In the particular case of a non-equilibrium transformation in which the temperature $T$ of the system is not varied, the latter expression can be written as~\cite{Jarzynski:1996ne}
\begin{equation}
\label{Jarzynski}
\left\langle \exp \left[ - \frac{W(\tin,\tfin)}{T} \right] \right\rangle = \exp \left( - \frac{\Ffin-\Fin}{T} \right),
\end{equation}
where $W(\tin,\tfin)$ denotes the total work done on the system during the transformation from $\tin$ to $\tfin$.

Before closing this section, we point out some important remarks.

First of all, as we discussed above, the evaluation of free-energy differences using Jarzynski's relation assumes $N \to \infty$ (with $\tin$ and $\tfin$ fixed and finite). In this limit, the time-discretization step $\tau$ becomes infinitesimally small, and from the continuity of $\lambda$ it follows that the $\pi_{\lambda(t_n)}$ and $\pi_{\lambda(t_{n+1})}$ distributions at all pairs of subsequent times become more and more overlapping. Correspondingly, in a Monte~Carlo simulation the aforementioned potential systematic effects related to the asymmetric r\^oles of $t_n$ and $t_{n+1}$ in the Markov evolution of a field configuration with transition probability $P_{\lambda(t_{n+1})}\left[ \phi(t_n) \to \phi(t_{n+1}) \right]$ depending on the parameter values at time $t=t_{n+1}$ (or, conversely, in the summands on the right-hand side of eq.~(\ref{discretized_exponential_work}), where the difference is evaluated by keeping the field configuration fixed to its value at $t=t_n$) are expected to vanish---an expectation which is indeed confirmed by our numerical results.

It is also instructive to discuss what happens in the opposite limit, i.e. for $N=1$. In this case, the calculation reduces to evaluating the exponential average of the work (in units of $T$) that is done on the system when its parameters are switched from $\lambda_{\mbox{\tiny{in}}}$ directly to $\lambda_{\mbox{\tiny{fin}}}$. In particular, according to the derivation above (in which the work done on the system is evaluated by computing the variation in energy on one of the configurations of the initial ensemble), one can realize that for $N=1$ the field configurations from the initial, equilibrium ensemble with parameters $\lambda_{\mbox{\tiny{in}}}$ are not ``evolved'' at all, and that the parameters of the system are instantaneously switched to their final values $\lambda_{\mbox{\tiny{fin}}}$ at $t=t_1=\tfin$: at this time, the work done on the system is calculated on the initial configuration, but then the configuration itself is not subject to any evolution, and, in particular, is not driven out of equilibrium at all. Interestingly, the exponential average of the work done on the system is exactly equal to $Z_{\lambda(\tfin)}/Z_{\lambda(\tin)}$ also in the $N=1$ case, as it was already pointed out in the first work in which Jarzynski's relation was derived~\cite{Jarzynski:1996ne}. In fact, the existence of a relation of this type has been known for a long time (see, e.g., ref.~\cite{Zwanzig:1954ht}), and does not involve any non-equilibrium evolution. From a lattice gauge theory point of view, in the $N=1$ case this calculation corresponds to computing the average value of the exponential of the difference in Euclidean action, that is induced by a change in the parameters characterizing the system; this average is performed in the starting ensemble, with partition function $Z_{\lambda(\tin)}$. Using a terminology that may be more familiar among lattice practitioners, this can be recognized as a reweighting technique~\cite{Ferrenberg:1988yz, Barbour:1997bh, Fodor:2001au}. Although this method to compute the free-energy difference of the initial and final ensembles is in principle exact, its practical applications in Monte~Carlo simulations of lattice QCD (which necessarily involve finite configuration samples) is of very limited computational efficiency, being affected by dramatically large uncertainties when the configuration probability distributions of the simulated ($\pi_{\lambda(\tin)}$) and target ($\pi_{\lambda(\tfin)}$) ensembles are poorly overlapping. Such \emph{overlap problem} becomes more severe when the probability distributions are more sharply peaked (which is the case for systems with a large number of degrees of freedom---including, in particular, lattice gauge theories defined on large and fine lattices) and/or more widely separated in configuration space, so that the simulation of the ensemble specified by the parameters $\lambda(\tin)$ samples only a very limited subset of the most likely configurations of the target ensemble.

What happens in the case when $N$ is finite and larger than one? In particular: in view of the previous observation, could one think that for finite $N>1$ the evaluation of the free-energy difference between the initial and the final ensemble by means of Jarzynski's relation is equivalent to a sequence of reweighting steps, at parameter values $\lambda(t_n)$, with $0 \le n < N$? The answer is no: a Monte~Carlo algorithm to compute the free-energy difference using Jarzynski's relation is crucially different from a combination of reweighting steps, because, in contrast to the former, the latter assumes that \emph{also} the field configurations at all later times $\phi(t_n)$, for $0 < n$, are drawn from equilibrium distributions. On the contrary, the sequence of field configurations produced during each trajectory in a numerical implementation of Jarzynski's relation are genuinely out of equilibrium: only the configurations at $t=\tin$ are drawn from an equilibrium distribution. As a consequence, there is no contradiction between the fact that the computation of the free-energy difference between two ensembles using Jarzynski's relation becomes exact only for \emph{infinite} $N$, and the fact that the same computation can also (at least in principle, i.e. neglecting the overlap problem mentioned above) be carried out exactly by reweighting the \emph{equilibrium} distributions defined at a \emph{finite} number of intermediate parameter values corresponding to $\lambda(t_n)$, with $0 \le n < N$. Similarly, there is no inconsistency in the fact that, using the algorithm based on Jarzynski's theorem for finite $N>1$, the results for $Z_{\lambda(\tfin)}/Z_{\lambda(\tin)}$ obtained from Monte~Carlo calculations in ``direct'' ($\lambda_{\mbox{\tiny{in}}} \to \lambda_{\mbox{\tiny{fin}}}$) and in ``reverse'' ($\lambda_{\mbox{\tiny{fin}}} \to \lambda_{\mbox{\tiny{in}}}$) evolutions of the system are not necessarily equal: they only have to agree in the large-$N$ limit---and, as our numerical results show, they do agree in that limit.

Note that these observations do not imply that a Monte~Carlo calculation of $Z_{\lambda(\tfin)}/Z_{\lambda(\tin)}$ using Jarzynski's relation, which requires $N$ to be large, is less efficient than one based on a combination of $N$ reweightings, which is exact for every value of $N$: on the contrary, the overhead of generating \emph{non-equilibrium} configurations at a larger number of intermediate values of the system parameters (whose computational cost grows like $O(N)$ and, for typical lattice gauge theory simulation algorithms, \emph{polynomially} in the number of degrees of freedom of the system), may be largely offset by the growth in statistics necessary for proper ensemble sampling in simulations using reweighting, which is \emph{exponential} in the number of degrees of freedom of the system~\cite{Gattringer:2016kco}. 

For a given physical system, in Monte~Carlo simulations based on Jarzynski's relation, the optimal choice of $N$ and of the number $\nr$ of ``trajectories'' in configuration space (or ``realizations'' of the non-equilibrium evolution of the system) over which the averages appearing on the left-hand-side of eqs.~(\ref{generalized_Jarzynski}) and (\ref{Jarzynski}) are evaluated, is the one minimizing the total computational cost, for a desired maximum level of uncertainty on the numerical results. In general, determining the optimal values of $N$ and $\nr$ is non-trivial, as they depend strongly on the system under consideration (and, often, on the details of the Monte~Carlo simulation). During the past few years, some aspects of this problem have been addressed in detail in various works: see refs.~\cite{Jarzynski:2006re, Pohorille:2010gp, Rohwer:2014co, YungerHalpern:2016hm} and references therein.

Finally, note that the derivation of Jarzynski's relation does not rely on any strong assumption about the nature of the system, and can be applied to every system with a Hamiltonian bounded from below. As such, it can be directly applied to statistical systems describing lattice gauge theories in Euclidean space. In the following, we present two applications of Jarzynski's relation in lattice gauge theory, first in the computation of the interface free energy in a gauge theory in three dimensions, and then in the calculation of the equation of state in $\SU(2)$ Yang--Mills theory in $3+1$ Euclidean dimensions.

\section{Benchmark study I: The interface free energy}
\label{sec:interface}

As a first benchmark study, we apply Jarzynski's relation eq.~(\ref{Jarzynski}) for a computation of the free energy associated with a fluctuating interface in a lattice gauge theory in three dimensions. As mentioned in section~\ref{sec:introduction}, interfaces have important experimental realizations in condensed-matter physics and in various other branches of science~\cite{Gelfand:1990fse, Privman:1992zv}. Moreover, they are also interesting for high-energy physics, as they can be related to the world-sheets spanned by flux tubes in confining gauge theories. Because of quantum fluctuations, the energy stored in a confining flux tube has a non-trivial dependence on its length~\cite{Luscher:1980fr, Luscher:1980ac}, which can be systematically studied in the framework of an effective theory~\cite{Aharony:2013ipa} and investigated numerically in lattice simulations (see refs.~\cite{Kuti:2005xg, Teper:2009uf, Panero:2012qx, Lucini:2012gg} for reviews). In particular, the effective action that describes the dynamics of flux tubes joining static color sources may include non-trivial terms associated with the boundaries of the string world-sheet~\cite{Aharony:2010cx, Aharony:2010db}. A possible way to disentangle the effect of these boundary contributions to the effective string action consists in studying closed string world-sheets, like those describing the evolution of a torelon (a flux loop winding around a spatial size of a finite system) over compactified Euclidean time: in that case the string world-sheet has the topology of a torus, and can be interpreted as a fluctuating interface. A closely related setup is relevant for the study of maximal 't~Hooft loops~\cite{'tHooft:1977hy, Srednicki:1980gb}.

The simplest lattice gauge theory, in which one can carry out a high-precision numerical Monte~Carlo study of interfaces, is the $\Z_2$ gauge model in three Euclidean dimensions, whose degrees of freedom are $\sigma_\mu(x)$ variables (taking values $\pm 1$) defined on the bonds between nearest-neighbor sites of a cubic lattice $\Lambda$ of spacing $a$. Following ref.~\cite{Caselle:2005vq}, we take the Euclidean action of the model to be the Wilson action~\cite{Wilson:1974sk}
\begin{equation}
S_{\mathbb{Z}_2} = - \betagauge \sum_{x \in \Lambda} \sum_{0 \le \mu < \nu \le 2} \sigma_\mu(x) \sigma_\nu(x+a\hat{\mu}) \sigma_\mu(x+a\hat{\nu}) \sigma_\nu(x)
\end{equation}
(where $\betagauge$ denotes the Wilson parameter for the $\Z_2$ gauge theory); it is trivial to verify that the model enjoys invariance under local $\Z_2$ transformations, that flip the sign of the $\sigma_\mu(x)$ link variables touching a given site. The partition function of the model reads
\begin{equation}
Z_{\mathbb{Z}_2} = \sum_{ \left\{ \sigma_\mu(x) = \pm 1 \right\} } \exp\left( -S_{\mathbb{Z}_2}\right).
\end{equation}
For small values of $\betagauge$ this model has a confining phase, which terminates at a second-order phase transition at $\betagauge = 0.76141346(6)$~\cite{Deng:2003wv}.

$Z_{\mathbb{Z}_2}$ can be exactly rewritten as the partition function of the three-dimensional Ising model~\cite{Kramers:1941kn, Wegner:1984qt}, whose degrees of freedom are $\Z_2$ variables $s_x$ defined on the sites of a dual cubic lattice $\widetilde{\Lambda}$, and whose Hamiltonian reads
\begin{equation}
H = - \beta \sum_{x \in \widetilde{\Lambda}} \sum_{0 \le \mu \le 2} J_{x,\mu} s_x s_{x+a\hat{\mu}},
\end{equation}
where $J_{x,\mu}=1$ corresponds to ferromagnetic couplings, while $J_{x,\mu}=-1$ would yield antiferromagnetic couplings, and $\beta$ and $\betagauge$ are related to each other by
\begin{equation}
\label{symmetric_beta_betagauge_relation}
\sinh(2\beta)\sinh(2\betagauge)=1.
\end{equation}
Note that, since $\sinh(2x)$ is a strictly increasing function, eq.~(\ref{symmetric_beta_betagauge_relation}) implies that the confining regime of the gauge theory (at small $\betagauge$) corresponds to the ordered phase of the Ising model (at large $\beta$). Eq.~(\ref{symmetric_beta_betagauge_relation}) can be rewritten as
\begin{equation}
\label{beta_betagauge_relation}
\beta=-\frac{1}{2} \ln \tanh \betagauge.
\end{equation}

Note that on a finite lattice, denoting the number of sites along the direction $\mu$ as $N_\mu$ and defining the site coordinates (in units of the lattice spacing) modulo $N_\mu$, one can impose periodic boundary conditions by setting all $J_{x,\mu}=1$, whereas antiperiodic boundary conditions in the direction $\mu$ can be imposed setting $J_{x,\mu}=-1$ only for the couplings between a spin in the first and a spin in the last lattice slice in direction $\mu$, i.e. $J_{x,\mu}=-1$ when $x_\mu/a=N_\mu-1$: in that case, a frustration is induced in the system, and an interface separating domains of opposite magnetization is formed. Finally, the choice $J_{x,\mu}=0$ for those bonds corresponds to decoupling the spins in the last lattice slice in direction $\mu$ from those in the first.

Thus, the ratio of the partition function with antiperiodic boundary conditions in one direction ($\Za$) over the one with periodic boundary conditions in all directions ($\Zp$) is directly related to the expectation value of an interface separating domains of different magnetizations. More precisely, if $N_0$ denotes the lattice size (in units of the lattice spacing $a$) in the direction in which antiperiodic boundary conditions are imposed, one can introduce a first definition of the interface free energy $F^{(1)}$ from
\begin{equation}
\label{f1_defining_relation}
\frac{\Za}{\Zp} = N_0 \exp\left( - F^{(1)} \right)
\end{equation}
(where the $N_0$ factor on the right-hand side accounts for the fact that the interface can be located anywhere along the direction in which antiperiodic boundary conditions are imposed), namely
\begin{equation}
\label{f1}
F^{(1)} = -\ln\left(\frac{\Za}{\Zp}\right) + \ln N_0.
\end{equation}
Note that here $F^{(1)}$ is defined as a dimensionless quantity. For a system of sufficiently large transverse cross-section (i.e. when the sizes $L_1$ and $L_2$ in the directions normal to the one in which antiperiodic boundary conditions are imposed are large), $F^{(1)}$ is expected to be proportional to $L_1 L_2$, with a positive proportionality coefficient. As a consequence, the expectation value of large interfaces is exponentially suppressed with their area, and one can assume that only one ``large'' interface (i.e. one extending through a whole cross-section of the system) is formed in the presence of antiperiodic boundary conditions---whereas no large interfaces are formed in the system with periodic boundary conditions. For a finite-size system, however, one can also consider the case of multiple large interfaces (in particular: an odd number of them for antiperiodic boundary conditions in one direction, and an even number of them for periodic boundary conditions). As discussed in ref.~\cite{Caselle:2007yc}, under the assumption that these interfaces are indistinguishable, dilute and non-interacting, one can derive an improved definition of the dimensionless interface free energy:
\begin{equation}
\label{f2}
F^{(2)} = -\ln  \arctanh \left( \frac{\Za}{\Zp} \right) + \ln N_0.
\end{equation}
Note that $F^{(2)}$ tends to $F^{(1)}$ when $\Za \ll \Zp$.

These definitions show that the dimensionless interface free energy can be evaluated in a numerical simulation, by computing the $\Za/\Zp$ ratio. As discussed above, $\Za$ and $\Zp$ can be interpreted as the partition functions of two systems that differ by the value of the $J_{x,\mu}$ couplings in one direction, that we have assumed to be the one labelled by $0$, on one slice (say, the one corresponding to $x_0=N_0-1$): $\Za$ is the partition function of the Ising spin system in which those couplings are set to $-1$ (while $J_{x,\mu}=1$ for $\mu \neq 0$ or for $x_0 \neq N_0-1$), whereas $\Zp$ is the partition function of the Ising spin system in which all couplings are ferromagnetic ($J_{x,\mu}=1$ for all $\mu$ and for all $x$). One can thus evaluate the $\Za/\Zp$ ratio by applying Jarzynski's relation eq.~(\ref{Jarzynski}), identifying the $J$ couplings on the $\mu=0$ bonds from the sites in the $x_0=N_0-1$ slice of the system as the $\lambda$ parameters to be varied as a function of Monte~Carlo time $t$. In particular, one can let those couplings vary linearly with time, interpolating from $J=1$ at $t=\tin$ to $J=-1$ at $t=\tfin$,
\begin{equation}
\label{J_evolution}
\lambda\left(\tin + n\tau \right) = J_{(N_0-1,x_1,x_2),0} \left( \tin + n\tau \right)= 1 - \frac{2n}{N}, \qquad \mbox{with}~\tau=\frac{\tfin-\tin}{N},
\end{equation}
for $~n \in \left\{ 0, 1, \dots , N \right\}$, or vice~versa. A similar application of Jarzynski's relation was used in the study of the Ising model in two dimensions~\cite{condmat0602580, Chatelain:2007ts, Hijar2007}. 

It is worth remarking that parallelization (as well as other standard algorithmic techniques for spin systems, like multi-spin coding) is straightforward to implement in a computation of the free energy based on Jarzynski's relation.

We carried out a set of Monte~Carlo calculations of the interface free energy using this method (with $N=10^6$ and averaging over $\nr=10^3$ realizations of the discretized non-equilibrium transformation), at the parameters used in the study reported in ref.~\cite{Caselle:2007yc}, finding perfect agreement with the results of that study. We also observed that the exponential work averages corresponding to a ``direct'' (from $\Zp$ to $\Za$) or a ``reverse'' (from $\Za$ to $\Zp$) parameter switch converge to the same results, and that the latter are independent of the $\lambda(t)$ parametrization at large $N$.

\begin{figure}[!htpb]
\centerline{\includegraphics[width=0.9\textwidth]{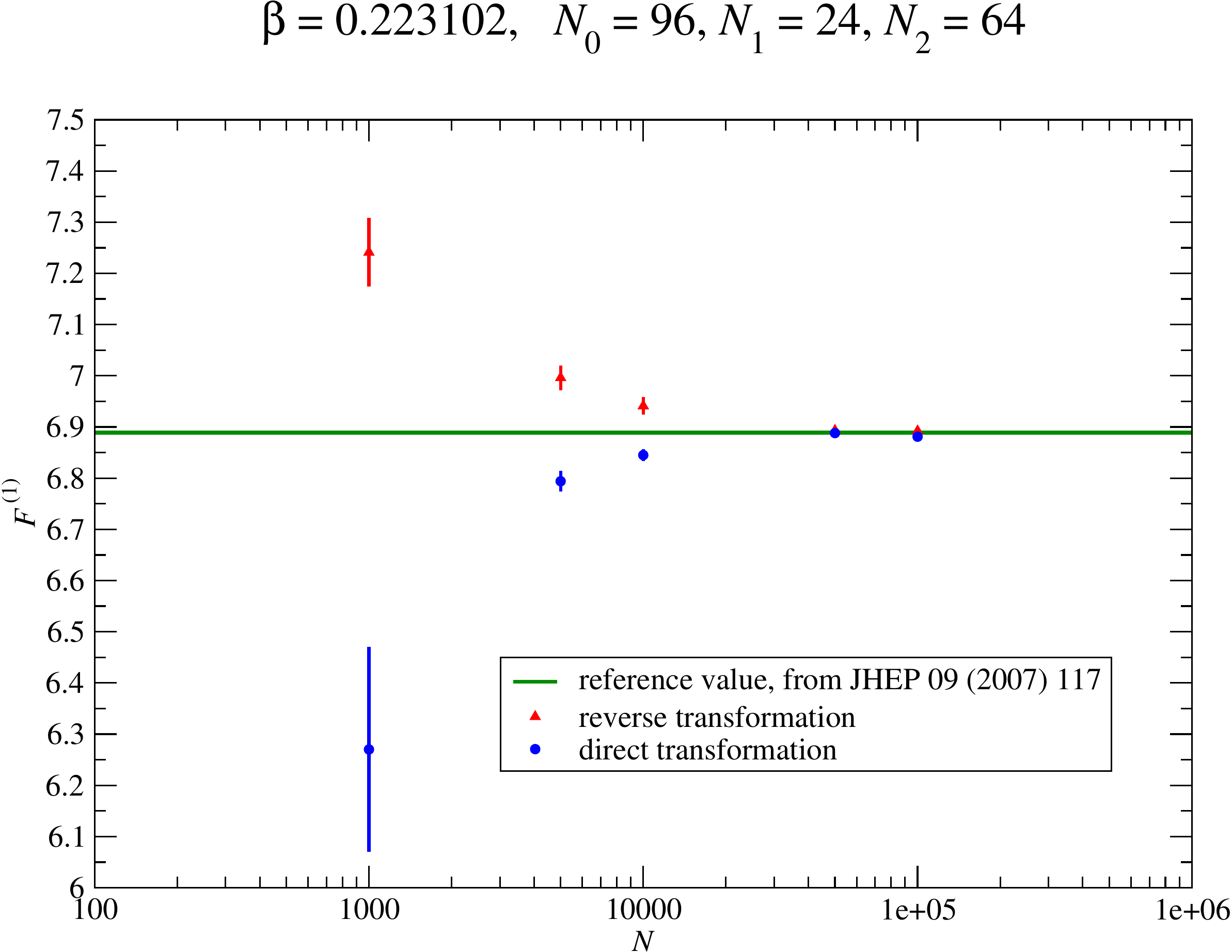}}
\caption{(Color online) Convergence of our results for the interface free energy---defined according to eq.~(\ref{f1})---obtained in direct (blue bullets) and reverse (red triangles) transformations from $\Zp$ to $\Za$ in Monte~Carlo simulations at $\beta=0.223102$ (corresponding to $\betagauge=0.758264$) on a lattice of sizes $L_0=96a$, $L_1=24a$, $L_2=64a$. The green band denotes the value of the interface free energy determined in ref.~\cite{Caselle:2007yc} for these values of the parameters, and with a different method. $N$ is the number of intervals used to discretize the temporal evolution of the parameter by which the boundary conditions of the system in direction $\mu=0$ are switched from periodic to antiperiodic, according to eq.~(\ref{J_evolution}).}
\label{fig:96_24_64}
\end{figure}

This can be clearly seen in tables~\ref{tab:96_48_64}, \ref{tab:96_24_64} and \ref{tab:96_32_32}, where we report results for the interface free energies in the three-dimensional $\Z_2$ gauge model at $\betagauge=0.758264$, obtained from Monte~Carlo simulations of the Ising model at $\beta=0.223102$. These tables show that the free-energy estimates obtained from a ``direct'' and a ``reverse'' realization of the non-equilibrium transformation from $\Zp$ to $\Za$ converge to the same value (which is consistent with earlier calculations carried out by different methods~\cite{Caselle:2007yc}), when the discretization of the parameter evolution involved in the non-equilibrium transformation is carried out with a sufficient number of points. The results obtained from simulations on a lattice of sizes $L_0=96a$, $L_1=24a$ and $L_2=64a$ are also displayed in fig.~\ref{fig:96_24_64}.

\begin{table}[!htb]
\centering
\begin{tabular}{|c||c|c||c|c|}
\hline
$N$ & $\nr$ & $F^{(1)}$, direct & $\nr$ & $F^{(1)}$, reverse \\
\hline
$10^{3}$         & $64 \cdot 320$ & $11.25(13)$  & $ 64 \cdot 80  $ & $12.19(11) $  \\
$5 \cdot 10^{3}$ & $64 \cdot 320$ & $11.23(8) $  & $ 64 \cdot 80  $ & $11.52(4)  $  \\   
$10^{4}$         & $64 \cdot 320$ & $11.33(5) $  & $ 64 \cdot 80  $ & $11.41(3)  $  \\
$5 \cdot 10^{4}$ & $64 \cdot 80 $ & $11.25(3) $  & $ 64 \cdot 80  $ & $11.33(2)  $  \\   
$10^{5}$         & $64 \cdot 80 $ & $11.29(2) $  & $ 64 \cdot 80  $ & $11.32(1)  $  \\
\hline
\end{tabular}
\caption{Results for the interface free energy defined in eq.~(\ref{f1}) from ``direct'' and ``reverse'' realizations of the non-equilibrium parameter transformation from periodic to antiperiodic boundary conditions in the $\mu=0$ direction, on a lattice with $N_0=96$, $N_1=48$, $N_2=64$, at $\beta=0.223102$ (i.e. at $\betagauge=0.758264$), and for a different number $N$ of intervals used to discretize the temporal evolution of $\lambda$. $\nr$ is the statistics used in the average over non-equilibrium processes. The interface free energy evaluated in ref.~\cite{Caselle:2007yc} for these parameters is $F^{(1)} = 11.3138(25)$.\label{tab:96_48_64}}
\end{table}
 
\begin{table}[!htb]
\centering
\begin{tabular}{|c||c|c||c|c|}
\hline
$N$ & $\nr$ & $F^{(1)}$, direct & $\nr$ & $F^{(1)}$, reverse \\
\hline
$10^{3}$         & $ 64 \cdot 320 $ & $  6.27(20)  $ & $ 64 \cdot 80  $ & $ 7.241(67) $ \\
$5 \cdot 10^{3}$ & $ 64 \cdot 320 $ & $  6.794(20) $ & $ 64 \cdot 80  $ & $ 6.996(24) $  \\   
$10^{4}$         & $ 64 \cdot 320 $ & $  6.845(12) $ & $ 64 \cdot 80  $ & $ 6.941(17) $ \\
$5 \cdot 10^{4}$ & $ 64 \cdot 80  $ & $  6.888(8)  $ & $ 64 \cdot 80  $ & $ 6.893(8)  $ \\
$ 10^{5}$        & $ 64 \cdot 80  $ & $  6.881(6)  $ & $ 64 \cdot 80  $ & $ 6.892(5)  $ \\
\hline
\end{tabular}
\caption{Same as in table~\ref{tab:96_48_64}, but for $N_0=96$, $N_1=24$, $N_2=64$. The reference value of the interface free energy at these parameters, taken from ref.~\cite{Caselle:2007yc}, is $F^{(1)}=6.8887(20)$. The results listed in this table are also plotted in fig.~\ref{fig:96_24_64}.\label{tab:96_24_64}}
\end{table}

\begin{table}[!htb]
\centering
\begin{tabular}{|c||c|c||c|c|}
\hline
$N$ & $\nr$ & $F^{(1)}$, direct & $\nr$ & $F^{(1)}$, reverse \\
\hline
$10^{3}$   & $64 \cdot 80$ & $5.68(7)  $ & $64 \cdot 80$ & $6.32(6)  $ \\
$10^{4}$   & $64 \cdot 80$ & $5.943(14)$ & $64 \cdot 80$ & $6.018(13)$ \\
$10^{5}$   & $64 \cdot 80$ & $5.979(4) $ & $64 \cdot 80$ & $5.982(4) $ \\
\hline
\end{tabular}
\caption{Same as in table~\ref{tab:96_48_64}, but for square interfaces with $N_0=96$, $N_1=N_2=32$.\label{tab:96_32_32}}
\end{table}

It is interesting to study how our determination of the interface free energy using Jarzynski's relation compares with those based on different techniques. A state-of-the-art example of the latter was reported in ref.~\cite{1401.7870}, using the so-called ``ensemble-switch'' method. Carrying out some numerical tests, we found that the computational efficiency of the two algorithms is similar. In general, the structure of the ensemble-switch algorithm makes it more demanding in terms of CPU time. On the other hand, we observed that the algorithm based on Jarzynski's relation typically leads to results affected by somewhat larger intrinsic fluctuations. An important difference between the ensemble-switch algorithm and ours is that, in contrast to the former, the latter can be parallelized in a more straightforward way. For large $N$, our algorithm has a similar efficiency as (and in some cases even outperforms) the ensemble-switch algorithm.

Having verified the convergence of the interface free energy estimates from our algorithm based Jarzynski's relation (for non-equilibrium transformations from one ensemble to the other, in both directions), we report some results from simulations on lattices of different sizes in tables~\ref{tab:96_xx_xx_beta0223102}, \ref{tab:96_xx_32_beta0223102}, \ref{tab:96_xx_48_beta0223102}, \ref{tab:96_xx_64_beta0223102}, \ref{tab:96_xx_80_beta0223102} and \ref{tab:96_xx_96_beta0223102} (from simulations at $\beta=0.223102$) and in table~\ref{tab:96_xx_64_beta0226102} (from simulations at $\beta=0.226102$).

\begin{table}[!htb]
\centering
\begin{tabular}{|c|c|c|}
\hline
$N_1=N_2$ & $F^{(1)} $ & $F^{(2)}$  \\
\hline
$18$ & $4.61969(21)$ & $3.9800(9)$  \\
$20$ & $4.68520(24)$ & $4.2252(6)$  \\
$22$ & $4.79156(32)$ & $4.4785(5)$  \\
$24$ & $4.94312(34)$ & $4.7412(5)$  \\
$28$ & $5.3850(5)$   & $5.3143(5)$  \\
$32$ & $5.9785(6)$   & $5.9583(6)$  \\
$36$ & $6.6849(7)$   & $6.6801(7)$  \\
$40$ & $7.4819(9)$   & $7.4809(9)$  \\
$44$ & $8.3653(12)$  & $8.3652(11)$ \\
$48$ & $9.3318(15)$  & $9.3318(13)$ \\
\hline
\end{tabular}
\caption{Interface free energies---evaluated according to eq.~(\ref{f1}) and to eq.~(\ref{f2}), and respectively reported in the second and in the third column---obtained from simulations at $\beta=0.223102$ (corresponding to $\betagauge=0.758264$) on lattices of square cross-section with $N_0=96$ and for different values of $N_1=N_2$ (first column). For a comparison, the corresponding values obtained in ref.~\cite{Caselle:2007yc} at the same $\beta$ and for $N_0=96$, $N_1=N_2=40$, are $F^{(1)}=F^{(2)}=7.481(1)$.}
\label{tab:96_xx_xx_beta0223102}
\end{table}

\begin{table}[!htb]
\centering
\begin{tabular}{|c|c|c|}
\hline
$N_1$ & $F^{(1)} $ & $F^{(2)}$  \\
\hline
$22$ & $5.1677(4)$ & $5.0520(5)$  \\ 
$24$ & $5.3257(5)$ & $5.2450(5)$  \\ 
$26$ & $5.4868(5)$ & $5.4301(6)$  \\ 
$28$ & $5.6503(6)$ & $5.6103(7)$  \\
\hline
\end{tabular}
\caption{Interface free energies evaluated according to eq.~(\ref{f1}) and to eq.~(\ref{f2}) (second and third column) obtained from simulations at $\beta=0.223102$ (i.e. for $\betagauge=0.758264$) on lattices with $N_0=96$ and rectangular cross-section, for different values of $N_1$ (first column) at $N_2=32$.}
\label{tab:96_xx_32_beta0223102}
\end{table}

\begin{table}[!htb]
\centering
\begin{tabular}{|c|c|c|}
\hline
$N_1$ & $F^{(1)} $ & $F^{(2)}$  \\
\hline
$22$ & $5.8304(5)$  & $5.8030(6)$  \\ 
$24$ & $6.1238(5)$  & $6.1088(6)$  \\ 
$28$ & $6.6984(8)$  & $6.6937(8)$  \\  
$32$ & $7.2520(9)$  & $7.2504(9)$  \\  
$36$ & $7.7876(11)$ & $7.7871(11)$ \\ 
$40$ & $8.3142(15)$ & $8.3140(15)$ \\ 
$44$ & $8.8255(16)$ & $8.8255(16)$ \\
\hline
\end{tabular}
\caption{Same as in table~\ref{tab:96_xx_32_beta0223102}, but from simulations on lattices with $N_0=96$ and $N_2=48$.}
\label{tab:96_xx_48_beta0223102}
\end{table}

\begin{table}[!htb]
\centering
\begin{tabular}{|c|c|c|}
\hline
$N_1$ & $F^{(1)} $ & $F^{(2)}$  \\
\hline
$18$ & $5.6068(6)$ & $5.5629(5)$ \\
$20$ & $6.0369(6)$ & $6.0190(6)$ \\
$22$ & $6.4676(7)$ & $6.4601(7)$ \\
$24$ & $6.8868(8)$ & $6.8836(8)$ \\
\hline
\end{tabular}
\caption{Same as in table~\ref{tab:96_xx_32_beta0223102}, but from simulations on lattices with $N_0=96$ and $N_2=64$. The results obtained in ref.~\cite{Caselle:2007yc} 
at this $\beta$ and for $N_0=96$, $N_1=24$, $N_2=64$, are $F^{(1)}=6.889(2)$ and $F^{(2)}=6.886(2)$.}
\label{tab:96_xx_64_beta0223102}
\end{table}

\begin{table}[!htb]
\centering
\begin{tabular}{|c|c|c|}
\hline
$N_1$ & $F^{(1)} $ & $F^{(2)}$  \\
\hline
$18$ &  $5.9318(6)$  &  $5.9095(6) $ \\
$20$ &  $6.5018(7)$  &  $6.4948(7) $ \\
$22$ &  $7.0654(8)$  &  $7.0631(8) $ \\
$24$ &  $7.6140(9)$  &  $7.6132(9) $ \\
$26$ &  $8.1412(11)$ &  $8.1410(11)$ \\
$28$ &  $8.6550(15)$ &  $8.6549(13)$ \\
$32$ &  $9.6341(17)$ &  $9.6341(17)$ \\
$36$ & $10.5758(20)$ & $10.5758(20)$ \\
\hline
\end{tabular}
\caption{Same as in table~\ref{tab:96_xx_32_beta0223102}, but from simulations on lattices with $N_0=96$ and $N_2=80$.}
\label{tab:96_xx_80_beta0223102}
\end{table}

\begin{table}[!htb]
\centering
\begin{tabular}{|c|c|c|}
\hline
$N_1$ & $F^{(1)} $ & $F^{(2)}$  \\
\hline
$18$ &  $6.2314(7)$  &  $6.2193(7)$  \\
$20$ &  $6.9412(8)$  &  $6.9383(8)$  \\
$22$ &  $7.6392(9)$  &  $7.6385(9)$  \\
$24$ &  $8.3137(12)$ &  $8.3135(10)$ \\
$26$ &  $8.9583(12)$ &  $8.9583(12)$ \\
$28$ &  $9.5840(17)$ &  $9.5840(17)$ \\
$32$ & $10.7834(20)$ & $10.7834(20)$ \\
\hline
\end{tabular}
\caption{Same as in table~\ref{tab:96_xx_32_beta0223102}, but from simulations on lattices with $N_0=N_2=96$.}
\label{tab:96_xx_96_beta0223102}
\end{table}

\begin{table}[!htb]
\centering
\begin{tabular}{|c|c|}
\hline
$N_1$ & $F^{(2)}$ \\
\hline
$18$ & $13.9858(24)$ \\
$20$ & $15.4881(29)$ \\
$22$ & $16.9667(31)$ \\
$24$ & $18.4127(34)$ \\
\hline
\end{tabular}
\caption{Interface free energy (second column), defined according to eq.~(\ref{f2}), from simulations at $\beta=0.226102$ (corresponding to $\betagauge=0.751805$) on lattices with $N_0=96$ and $N_2=64$, for various values of $N_1$ (first column). The result reported in ref.~\cite{Caselle:2007yc} at this $\beta$, for $N_0=96$, $N_1=24$, and $N_2=64$, is $F^{(2)}=18.4131(26)$.}
\label{tab:96_xx_64_beta0226102}
\end{table}

These high-precision results can be directly compared with an effective theory, describing the  transverse fluctuations of the interface at low energies. In direct analogy with the effective description of the world-sheets associated with fluctuating, string-like flux tubes in confining gauge theories~\cite{Kuti:2005xg}, or with solitonic strings in Abelian Higgs models~\cite{Abrikosov:1956sx, Nielsen:1973cs}, this effective theory must be consistent with the Lorentz--Poincar\'e symmetries of the space in which the interface is defined~\cite{Aharony:2009gg} (see also ref.~\cite{Meyer:2006qx}). This condition puts strong constraints on the coefficients of the possible terms appearing in the effective action of the theory, making the latter very predictive: in particular, one finds that, on sufficiently long distances, the dynamics can be approximated very well by assuming that the possible ``configurations'' of the fluctuating interface occur with a Boltzmann weight $\exp(-\Seff)$, in which $\Seff$ is proportional to the \emph{area} of the interface itself, i.e. the effective action tends to the Nambu--Got\={o} action~\cite{Goto:1971ce, Nambu:1974zg}
\begin{equation}
\label{Nambu-Goto_action}
\Seff \simeq \sigma \int \dd^2 \xi \sqrt{\det g_{\alpha\beta}},
\end{equation}
where $\xi$ are coordinates parametrizing the interface surface, while $g_{\alpha\beta}$ is the metric induced by the embedding of the interface in the target space, while $\sigma$ can be thought of as the tension associated with the interface, in the classical limit. As discussed in ref.~\cite{Aharony:2009gg}, the actual form of the effective action deviates from the expression on the right-hand side of eq.~(\ref{Nambu-Goto_action}) by terms which, for the problem of interest (a closed interface of linear size denoted as $L$, in a three-dimensional space) scale at least with the seventh inverse power of $L$.

The partition function associated with an interface described by the Nambu--Got\={o} effective action in eq.~(\ref{Nambu-Goto_action}) has been calculated analytically in ref.~\cite{Billo:2006zg}: for a system in $D$ spacetime dimensions, this computation predicts
\begin{equation}
\label{interface_partition_function}
\frac{\Za}{\Zp} = 2 \mathcal{C} \left( \frac{\sigma}{2\pi} \right)^{\frac{D-2}{2}}\, V_{\mbox{\tiny{T}}} \, \sqrt{\sigma L_1 L_2 u}
\sum_{k=0}^\infty  \sum_{k'=0}^\infty c_k c_{k'} 
\left(\frac{\mathcal{E}_{k,k'}}{u}\right)^{\frac{D-1}{2}}\, K_{\frac{D-1}{2}} \left(\sigma L_1 L_2 \mathcal{E}_{k,k'}\right) = \mathcal{C} \mathcal{I},
\end{equation}
where $u=L_2/L_1$, $V_{\mbox{\tiny{T}}}$ denotes the ``volume'' of the system along the dimensions transverse to the interface (so $V_{\mbox{\tiny{T}}}=L_0$ in our case), $K_\nu(z)$ denotes the modified Bessel function of the second kind of order $\nu$ and argument $z$, while $c_k$ and $c_{k'}$ are coefficients appearing in the expansion of an inverse power of Dedekind's $\eta$ function:
\begin{equation}
\frac{1}{\eta\left( i u \right)^{D-2}} = \sum_{k=0}^{\infty}c_k q^{k-\frac{D-2}{24}},\qquad \mbox{with}\;\; q=\exp\left( -2\pi u \right)
\end{equation}
(so that, for the $D=3$ case, $c_k$ equals the number of partitions of $k$) and
\begin{equation}
\label{string_energies}
\mathcal{E}_{k,k'}  = \sqrt{ 1 + \frac{4\pi\, u}{\sigma  L_1 L_2 }\left(k+k'-\frac{D-2}{12}\right) + \left[\frac{2\pi u (k-k')}{\sigma L_1 L_2 }\right]^2}.
\end{equation}
Finally, $\mathcal{C}$ is an undetermined, non-universal multiplicative constant, which is not predicted by the effective bosonic-string model (and, following the notations of ref.~\cite{Billo:2006zg}, in the last term of eq.~(\ref{interface_partition_function}) we define the ratio of $\Za/\Zp$ over $\mathcal{C}$ as $\mathcal{I}$). Similarly, the model does not predict the value of the multiplicative constant involved in the partition function associated with one (or more) static color source(s): see refs.~\cite{Mykkanen:2012ri, Mykkanen:2012dv} for a discussion. These aspects are related to the fact that the bosonic-string model is a low-energy effective theory, which cannot capture non-universal terms whose origin involves ultraviolet dynamics.

The accuracy of this effective theory depends on the dimensionless parameter $1/(\sigma L_1 L_2)$: when this parameter is small, the bosonic-string model is expected to provide a good description of the interface free energy.

Given that the algorithm based on Jarzynski's relation allows one to reach high numerical precision, it is particularly interesting to compare our results for the interface free energy with the predictions from the effective string model, trying to identify deviations from the  terms predicted by a Nambu--Got\={o} action. To this purpose, we analyzed the results obtained at $\beta=0.223102$ expressing all dimensionful quantities in units of the interface tension $\sigma$: as determined in ref.~\cite{Caselle:2007yc}, at this value of $\beta$, one has $\sigma a^2 = 0.0026083(6)(7)$. Note that this implies that the lattice spacing is quite small, so discretization effects should be under control.

The first step in this analysis consists in subtracting the Nambu--Got\={o} prediction for the free energy, obtained from the logarithm of the r.h.s. of the first equality in eq.~(\ref{interface_partition_function}), from our data. Since  eq.~(\ref{interface_partition_function}) predicts the $\Za/\Zp$ ratio only up to the undetermined multiplicative constant $\mathcal{C}$, it predicts the interface free energy only up to an additive term $q=-\ln\mathcal{C}$. Like $\mathcal{C}$, $q$ depends only on the ultraviolet details of the theory, namely it can depend on the lattice spacing $a$ (or, equivalently, on $\beta$), but not on the lattice sizes. At each $\beta$, the value of $q$ can be fixed, by observing that the corrections to the Nambu--Got\={o} prediction are expected to become negligible for sufficiently large interfaces, i.e. for $\sigma L_1 L_2 \gg 1$. To this purpose, for each combination of values of $\sigma a^2$ and lattice sizes, we define our numerical estimate of the free energy ($F_{\mbox{\tiny{num}}}$) according to eq.~(\ref{f1}), using the results of our Monte~Carlo simulations for the $\Za/\Zp$ ratio. Then, for the same combination of $\sigma a^2$ and lattice sizes, we compute the quantity $\mathcal{I}$ appearing in eq.~(\ref{interface_partition_function}), and we define a quantity (denoted as $F_{\mathcal{I}}$) using $\mathcal{I}$ in place of the $\Za/\Zp$ ratio in eq.~(\ref{f2}). It is easy to see that $q$ can be obtained from
\begin{equation}
\label{q_definition}
q = \lim_{\sigma L_1 L_2 \to \infty} \left( F_{\mbox{\tiny{num}}} - F_{\mathcal{I}} \right)
\end{equation}
(note that, when $ \sigma L_1 L_2$ is large, the $\Za/\Zp$ ratio tends to zero, and, as discussed above, the free-energy definitions given by eqs.~(\ref{f1}) and (\ref{f2}) become equivalent). For every value of $\sigma a^2$, when $\sigma L_1 L_2$ becomes large our results for the $F_{\mbox{\tiny{num}}}-F_{\mathcal{I}}$ difference tend, indeed, to a constant, and can be fitted to $q=0.9168(5)$.

Then, we study the deviations of our Monte~Carlo results from the Nambu--Got\={o} predictions (for each value of $\beta$, and for each combination of lattice sizes), by defining the difference
\begin{equation}
\label{y_definition}
y = F_{\mbox{\tiny{num}}} - F_{\mathcal{I}} - q.
\end{equation}
This quantity depends on the interface sizes $L_1$ and $L_2$, and encodes the contributions to the free energy from terms appearing in the effective string action, that do not arise from in a low-energy expansion of the Nambu--Got\={o} action (and/or from possible systematic effects, related for example to the finiteness of the lattice spacing; however, previous studies indicate that the latter should be very modest for $\beta$ values in the range under consideration here---see, for example, ref.~\cite{Caselle:2005vq} and references therein). For each set at fixed $L_2 > 32a$, these data can be successfully fitted to the form expected for the leading and next-to-leading corrections to the Nambu--Got\={o} model, which, according to the discussion in ref.~\cite{Aharony:2013ipa}, scale with the seventh and with the ninth inverse power of the $L_1$ length scale:\footnote{The fact that, in three spacetime dimensions, the leading correction to the Nambu--Got\={o} model scales at least with the seventh inverse power of the length scale of the system has been recently observed also in lattice simulations of $\SU(N)$ gauge theories~\cite{Athenodorou:2016kpd}.}
\begin{equation}
\label{corrections_to_NG}
y = \frac{1}{\left( L_1 \sqrt{\sigma} \right)^{7}} \left[ k_{-7} + \frac{k_{-9}}{\left( L_1 \sqrt{\sigma} \right)^{2}} \right].
\end{equation}
Note that, for an interface in $D=3$ dimensions, the effective-string arguments indicate that additional subleading corrections, not included in the expression on the right-hand side of eq.~(\ref{corrections_to_NG}), are expected to be $O\left( (L_1 \sqrt{\sigma} )^{-11} \right)$, i.e. to be suppressed by at least one further factor of $1/(L_1^2 \sigma)$. The results of these fits are reported in table~\ref{tab:corrections_to_NG_fit}, where $\redchisq$ denotes the reduced $\chi^2$ obtained in the fit, i.e. the ratio of the $\chi^2$ over the number of degrees of freedom.

\begin{table}[!htb]
\centering
\begin{tabular}{|c|c|c|c|}
\hline
$N_2$ & $k_{-7}$ & $k_{-9}$ & $\redchisq$ \\
\hline
$48$ & $0.389(1)$ & $0.03(3)$  & $1.09$ \\
$64$ & $0.432(2)$ & $0.22(3)$  & $1.06$ \\
$80$ & $0.593(2)$ & $0.25(3)$  & $1.47$ \\
$96$ & $0.650(5)$ & $0.410(7)$ & $0.07$ \\
\hline
\end{tabular}
\caption{Results of our fits of the difference between our numerical results for the interface free energy and the corresponding Nambu--Got\={o} prediction, as defined in the text, to eq.~(\ref{corrections_to_NG}).}
\label{tab:corrections_to_NG_fit}
\end{table}

We also observed that a single inverse-power (of $L_1 \sqrt{\sigma}$) correction is not sufficient to describe our data. When we tried to set $k_{-9}$ to zero, leaving $k_{-7}$ as the only parameter to be fitted in eq.~(\ref{corrections_to_NG}), we always obtained values of the $\chi^2$ per degree of freedom much larger than $1$ (e.g. $\redchisq \simeq 24$ for $N_2=80$, and $\redchisq \simeq 61.5$ for the $N_2=96$ case), indicating that a term of order $\left( L_1 \sqrt{\sigma} \right)^{-7}$ alone does not fit our numerical results for $y$. In addition, we also observed that, if $k_{-9}$ is set to zero, but the exponent of $L_1 \sqrt{\sigma}$ for the other term (besides its coefficient) is treated as a fit parameter, i.e. if we make the \emph{Ansatz}
\begin{equation}
\label{single-power_correction}
y = \frac{k}{\left( L_1 \sqrt{\sigma} \right)^{\alpha}}
\end{equation}
with $k$ and $\alpha$ as fit parameters, the fits yield values of $\alpha$ that are incompatible across the data sets corresponding to different $N_2$, and that increase with $N_2$, ranging from $7.10(8)$ (for $N_2=48$), to $7.54(7)$ (for $N_2=64$), to $7.44(5)$ (for the data set at $N_2=80$), to $7.60(2)$ (for $N_2=96$). While the value of $\alpha$ obtained from the data set at $N_2=48$ may be compatible with $7$, the others, clearly, are not: the results at $N_2=64$ and $N_2=80$ may be compatible with a half-integer exponent $15/2$ (for which, however, there is no theoretical justification), but this is not the case for those at $N_2=96$. We also observe that the values of the reduced $\chi^2$ for some of these fits are significantly larger than $1$ (for example, $\redchisq$ is around $1.7$ for the data sets corresponding to $N_2=48$ and to $N_2=80$). This led us to conclude that our numerical results for the deviations from the Nambu--Got\={o} model cannot be fitted to a functional form including a correction given by a single inverse power of $L_1 \sqrt{\sigma}$ of the form given in eq.~(\ref{single-power_correction}).

These results support the expectations from the effective string model discussed in ref.~\cite{Aharony:2013ipa}; however, a puzzle remains: the values of $k_{-7}$ and $k_{-9}$ extracted from the fits have a residual dependence on $L_2$, whose origin is not clear. This could indicate that, as already pointed out in ref.~\cite{Caselle:2010pf}, the effective action describing the low-energy dynamics of this gauge theory includes additional terms. One possible such term could be the one describing the string ``stiffness''~\cite{Polyakov:1986cs, Kleinert:1986bk, Braaten:1986bz, German:1989vk, Klassen:1990dx}. We postpone a detailed analysis of this problem to a future, dedicated study.

We conclude this section with an important remark: even though we have calculated the interface free energy of the $\Z_2$ lattice gauge in three dimensions by mapping it to the Ising model, this was not a necessary condition for the application of Jarzynski's relation. An explicit example of application of Jarzynski's relation directly in a lattice gauge theory is presented in the following section~\ref{sec:equation_of_state}.

\section{Benchmark study II: The equation of state}
\label{sec:equation_of_state}

As another example of application of Jarzynski's relation eq.~(\ref{Jarzynski}) in lattice gauge theory, we discuss the calculation of the pressure in $\SU(2)$ Yang--Mills theory in $D=4$ spacetime dimensions. As is well-known, this gauge theory has a second-order deconfinement phase transition at a finite critical temperature $T_c$~\cite{Fingberg:1992ju, Engels:1994xj, Lucini:2005vg}, which, when expressed in physical units, is approximately $300$~MeV~\cite{Teper:1998kw, Lucini:2002ku, Lucini:2003zr, Lucini:2005vg}. As usual, the main quantities describing the thermal equilibrium properties of this theory are the pressure ($p$), the energy density ($\epsilon=E/V$) and the entropy density ($s=S/V$); these observables are related to each other by standard thermodynamic identities:
\begin{equation}
\epsilon = (D-1)p + \Delta, \qquad s =\frac{Dp + \Delta}{T},
\end{equation}
where $\Delta$ is the trace of the energy-momentum tensor, which, in turn, satisfies the relation
\begin{equation}
\Delta = T^{D+1} \frac{\partial}{\partial T} \left( \frac{p}{T^D} \right).
\end{equation}
As we mentioned in section~\ref{sec:introduction}, in the thermodynamic limit $V\to \infty$, the pressure equals minus the free-energy density, $p=-f=-F/V$, and this opens up the possibility to evaluate it using Jarzynski's relation. More precisely, we focus our attention on determining how the pressure depends on the temperature in the confining phase, i.e. at temperatures $T< \Tc$, assuming that the pressure vanishes for $T=0$. As it was recently shown in ref.~\cite{Caselle:2015tza}, the equilibrium-thermodynamics properties in the confining phase of this theory can be modelled very well in terms of a gas of free glueballs, using the masses of the lightest states known from previous lattice studies~\cite{Teper:1998kw} and assuming that the spectral density of heavier states has an exponential form~\cite{Hagedorn:1965st} (see also refs.~\cite{Buisseret:2011fq, Arriola:2014bfa} for discussions on related topics, and ref.~\cite{Caselle:2011fy} for an analogous lattice study in $2+1$ dimensions). Similar results have also been obtained in lattice studies of $\SU(3)$ Yang--Mills theory~\cite{Meyer:2009tq, Borsanyi:2012ve} and may be of direct phenomenological relevance even for real-world QCD~\cite{Stoecker:2015zea, Stocker:2015nka}.

It is worth remarking that the lattice determination of the equation of state in the confining phase of $\SU(N)$ Yang--Mills theory is not a computationally trivial problem: at low temperatures, the thermodynamic quantities mentioned above take values, that are significantly smaller than in the deconfined phase ($T > \Tc$). In the hadron-gas picture, the exponential suppression of these thermodynamic quantities for $T \ll \Tc$ is a direct consequence of confinement, i.e. of the existence of a finite mass gap---a relatively large one: when converted to physical units, the mass of the lightest glueball is around $1.6$~GeV for both $\SU(2)$~\cite{Teper:1998kw, Lucini:2001ej, Lucini:2004my} and $\SU(3)$~\cite{Morningstar:1999rf} Yang--Mills theories. 

Here, we focus on $\SU(2)$ Yang--Mills theory in four spacetime dimensions, and, following the notations of ref.~\cite{Caselle:2015tza}, we discretize it on a isotropic hypercubic lattice of spacing $a$ by introducing Wilson's gauge action~\cite{Wilson:1974sk}:
\begin{equation}
\label{Wilson_action}
S_{\SU(2)} = -\frac{2}{g^2} \sum_{x \in \Lambda} \sum_{0 \le \mu < \nu \le 3} \Tr U_{\mu\nu} (x),
\end{equation}
where $g$ is the coupling, related to $\betagauge$ via $\betagauge=4/g^2$, and $U_{\mu\nu} (x)$ denotes the plaquette from the site $x$ and lying in the oriented $(\mu,\nu)$ plane:
\begin{equation}
\label{plaquette}
U_{\mu\nu} (x) = U_\mu (x) U_\nu \left(x+a\hat{\mu}\right) U_{\mu}^\dagger \left(x+a\hat{\nu}\right) U_{\nu}^\dagger (x),
\end{equation}
where $\hat{\mu}$ and $\hat{\nu}$ denote unit vectors in the positive $\mu$ and $\nu$ directions, respectively. In the following, we assume that the compactified Euclidean-time direction is the $\mu=0$ direction, so that $T=1/(aN_0)$, while we take the lattice sizes in the three other directions to be equal ($N_1=N_2=N_3$, that we denote as $N_s$) and sufficiently large, to avoid finite-volume effects. Note that, in order to control the temperature of the system, we used the relation between $a$ and the inverse coupling $\betagauge$ determined in ref.~\cite{Caselle:2015tza}, and discussed in the next paragraph, so that we were able to change the temperature $T$ simply by varying $\betagauge$ at fixed $N_0$. We denote the normalized expectation value of the average of the trace of the plaquette at a generic temperature $T$ as $\langle U_{\Box}\rangle_T$: this quantity is averaged over all sites of the lattice and over all of the distinct $(\mu,\nu)$ planes, and is normalized to $1$ by dividing the trace by the number of color charges, i.e. by $2$ for the $\SU(2)$ gauge theory.

In order to ``set the scale'' of the lattice theory (i.e. to define a physical value for the lattice spacing $a$, as a function of $\betagauge$), we use the same non-perturbative procedure as in ref.~\cite{Caselle:2015tza}, based on the determination of the value of the force between static fundamental color sources at asymptotically large distances (i.e. the string tension of the theory) in lattice units, $\sigma a^2$: in the $2.25 \le \betagauge \le 2.6$ range, the relation between $a$ and $\betagauge$ is parametrized as
\begin{equation}
\label{betaform}
\ln \left( \sigma a^2 \right) = \sum_{j=0}^{3} h_j \left( \betagauge - \betagauge^{\mbox{\tiny{ref}}} \right)^j,
\end{equation}
where $\betagauge^{\mbox{\tiny{ref}}}=2.4$, while $h_0 = -2.68 $, $h_1 = -6.82 $, $h_2 = -1.90 $ and $h_3 = 9.96$. In addition, we mention that, for this gauge theory, the value of the ratio of the deconfinement critical temperature over the square root of the string tension is $T_c/\sqrt{\sigma} = 0.7091(36)$~\cite{Lucini:2003zr}.

A popular technique to compute the pressure $p$ (as a function of the temperature $T$, and with respect to the pressure at a conventional reference temperature: usually one defines $p$ as the difference with respect to the value it takes at $T=0$, which can be assumed to vanish) in lattice gauge theory is the integral method introduced in ref.~\cite{Engels:1990vr}. Here we describe it for the pure Yang--Mills theory. The method is based on the fact that, as we mentioned above, in the thermodynamic limit the pressure equals minus the free energy density; in turn, this quantity is proportional to the logarithm of the partition function, which can be computed by integrating its derivative with respect to the Wilson parameter $\betagauge$. At $T=0$ the pressure is vanishing, hence one could think of defining it as
\begin{equation}
\label{unphysical_pressure}
p^{\mbox{\tiny{unphys}}} = - f = \frac{T}{V} \ln Z =  \frac{1}{a^4 N_0 N_s^3} \int_{\betagauge^{(0)}}^{\betagauge^{(T)}} \dd \betagauge \frac{\partial \ln Z}{\partial \betagauge},
\end{equation}
where the upper integration extremum $\betagauge^{(T)}$ is the value of Wilson's parameter at which the lattice spacing $a$ equals $1 / \left( N_0 T \right)$, while the lower integration extremum $\betagauge^{(0)}$ is a value of Wilson's parameter, corresponding to a lattice spacing $a^{(0)}$ sufficiently large, so that the temperature $1/\left(a^{(0)}N_0\right)$ is close to zero. Using the fact that the logarithmic derivative of $Z$ with respect to $\betagauge$ equals the plaquette expectation value times the number of plaquettes (which is $6 N_0 N_s^3$), eq.~(\ref{unphysical_pressure}) reduces to
\begin{equation}
\label{unphysical_pressure_bis}
p^{\mbox{\tiny{unphys}}} = \frac{6}{a^4} \int_{\betagauge^{(0)}}^{\betagauge^{(T)}} \dd \betagauge \langle U_{\Box} \rangle_{\mathcal{T}(\betagauge)},
\end{equation}
where $\mathcal{T}(\betagauge)=1/\left[N_0 a(\betagauge)\right]$ is the temperature of the theory defined on a lattice with $N_0$ sites along the Euclidean-time direction and at Wilson parameter $\betagauge$, corresponding to a lattice spacing $a(\betagauge)$.

However, a definition of the pressure according to eq.~(\ref{unphysical_pressure_bis}) is actually unphysical (whence the $^{\mbox{\tiny{unphys}}}$ superscript), because it diverges in the continuum limit. This is easy to see, by inspection of eq.~(\ref{unphysical_pressure_bis}): in the $a \to 0$ limit, the integrand appearing on the right-hand side is a quantity that remains $O(1)$ in the whole integration domain, and the integral is multiplied by the divergent factor $6/a^4$. This unphysical ultraviolet divergence can be removed by subtracting the plaquette expectation value at $T=0$ (and at the same $\betagauge$) from the integrand on the right-hand side of eq.~(\ref{unphysical_pressure_bis}). This leads to the correct physical definition of the pressure according to the integral method:
\begin{equation}
\label{integral_method}
p = \frac{6}{a^4} \int_{\betagauge^{(0)}}^{\betagauge^{(T)}} \dd \betagauge \left[ \langle U_{\Box} \rangle_{\mathcal{T}(\betagauge)} - \langle U_{\Box} \rangle_0 \right],
\end{equation}
where $\langle U_{\Box}\rangle_0$ is evaluated from simulations on a symmetric lattice of sizes $N_s^4$ at the same value of $\betagauge$ (i.e. at the same lattice spacing) as $\langle U_{\Box}\rangle_{\mathcal{T}(\betagauge)}$.

Accordingly, the dimensionless $p(T)/T^4$ ratio can be evaluated as
\begin{equation}
\label{lattice_pressure}
\frac{p}{T^4} = 6 N_0^4 \int_{\betagauge^{(0)}}^{\betagauge^{(T)}} \dd \betagauge \left[ \langle U_{\Box}\rangle_{\mathcal{T}(\betagauge)} - \langle U_{\Box}\rangle_0 \right].
\end{equation}

Thus, the integral method reduces the computation of the pressure to an integration of differences between the plaquette expectation values at finite ($\mathcal{T}$) and at zero temperature. Such integration can be carried out numerically (e.g. using the trapezoid rule, or some of the methods listed in ref.~\cite[Appendix]{Caselle:2007yc}), once the $\langle U_{\Box}\rangle_{\mathcal{T}(\betagauge)} - \langle U_{\Box}\rangle_0$ differences are known to sufficient precision, and at a large enough number of values of the Wilson parameter in the $\left[\betagauge^{(0)},\betagauge^{(T)}\right]$ interval. 

Note that eq.~(\ref{lattice_pressure}) reveals a potentially challenging aspect of the lattice determination of the equation of state obtained with the integral method, in the extrapolation to the continuum limit. The pressure, in units of the fourth power of the temperature, is evaluated as the product of $6N_0^4$ times the integral of a difference in plaquette expectation values. For a fixed temperature $T$, the number of lattice sites in the Euclidean-time direction $N_0=1/(aT)$ becomes large in the continuum limit $a \to 0$, and, since the $p(T)/T^4$ ratio tends to a finite constant (its physical value) in that limit, while the integration range remains finite, this means that at the same time the $\langle U_{\Box}\rangle_{\mathcal{T}} - \langle U_{\Box}\rangle_0$ differences must necessarily become small, scaling like $a^4$. This implies that, in a numerical simulation, both $\langle U_{\Box}\rangle_{\mathcal{T}}$ and $\langle U_{\Box}\rangle_0$ have to be determined with relative statistical uncertainties $O(a^4)$, which requires a computational effort scaling (at least) like $O(N_0^8)$. 

This significant computational cost provides a motivation to use Jarzynski's relation for the numerical computation of the pressure; in this case, $\lambda$ can be taken to be Wilson's parameter, which is let vary from $\betagauge^{(0)}$ at $t=\tin$, to $\betagauge^{(T)}$ at $t=\tfin$. A potential advantage of determining the equation of state this way, is that, in contrast to the standard implementation of the integral method described above, it would not require complete equilibration of the system at all intermediate values of $\betagauge$, and, hence, could reduce the computational cost of the calculation, at least by a factor. While there is no obvious reason to expect that the computational costs of an algorithm based on Jarzynski's relation could scale with a lower power of $N_0$ when the continuum limit is approached, its intrinsic non-equilibrium nature suggests that it could nevertheless be significantly cheaper than a standard algorithm to implement eq.~(\ref{lattice_pressure}), because it would dramatically reduce the costs associated with thermalization (only the configurations in the starting ensemble need to be equilibrated).

We computed the pressure of the theory at different temperatures $0 < T < \Tc$, using the method based on Jarzynski's relation eq.~(\ref{Jarzynski}), assuming the equality of the pressure and minus the density of free energy, and using the ``physical'' definition of the pressure, consistent with eq.~(\ref{integral_method}), in which the unphysical ultraviolet divergences are subtracted. For later convenience, in order to allow a direct comparison with the results obtained in ref.~\cite{Caselle:2015tza}, in which this subtraction was carried out using lattices of sizes $\widetilde{N}^4$ (where $\widetilde{N}$ can be different from $N_s$, but it must be sufficiently large to enforce that the temperature is close to zero, and to avoid systematic uncertainties due to finite-volume effects) instead of $N_s^4$, we include this slight generalization of the divergence-subtraction procedure discussed above in the present discussion. Moreover, we also relax the assumption that the starting temperature $T_0=1/\left[ a\left(\betagauge^{(0)}\right) N_0\right]$ is close to zero, and that $p(T_0)$ vanishes.

As already mentioned in section~\ref{sec:Jarzynski}, we interpreted the differences appearing on the right-hand side of eq.~(\ref{discretized_exponential_work}) as differences in the Euclidean action of the lattice theory---i.e. as differences in the Wilson action defined in eq.~(\ref{Wilson_action})---when the $\betagauge$ parameter is varied. This leads to the following formula for the determination of $p/T^4$:
\begin{equation}
\label{lattice_pressure_Jarzynski}
\frac{p(T)}{T^4} = \frac{p(T_0)}{T_0^4} + \left( \frac{N_0}{N_s} \right)^3 \ln \frac{ \langle \exp \left[ - \Delta S_{\SU(2)} (\tin,\tfin)_{N_0 \times N_s^3} \right] \rangle }{ \langle \exp \left[ - \Delta S_{\SU(2)} (\tin,\tfin)_{\widetilde{N}^4} \right] \rangle^{\gamma} } ;
\end{equation}
on the right-hand side of this expression, $\Delta S_{\SU(2)} (\tin,\tfin)_{N_0 \times N_s^3}$ is the total variation in Wilson action calculated on a lattice of sizes $N_0 \times N_s^3$ during a non-equilibrium trajectory starting from a configuration of the initial, equilibrium ensemble with Wilson parameter $\betagauge^{(0)}$ realized at $t=\tin$, to a final configuration, obtained driving the system out of equilibrium until $\betagauge$ reaches its value $\betagauge^{(T)}$ at $t=\tfin$, and the $\langle \dots \rangle$ notation indicates averaging over $\nr$ such trajectories, as discussed in section~\ref{sec:Jarzynski}. Similarly, $\Delta S_{\SU(2)} (\tin,\tfin)_{\widetilde{N}^4}$ denotes an analogous total variation in Wilson action, but evaluated on a lattice of sizes $\widetilde{N}^4$, while the exponent $\gamma = \left( N_0 \times N_s^3 \right) / \widetilde{N}^4$ is the ratio of the lattice hypervolumes.

\begin{figure}[!htpb]
\centerline{\includegraphics[width=0.9\textwidth]{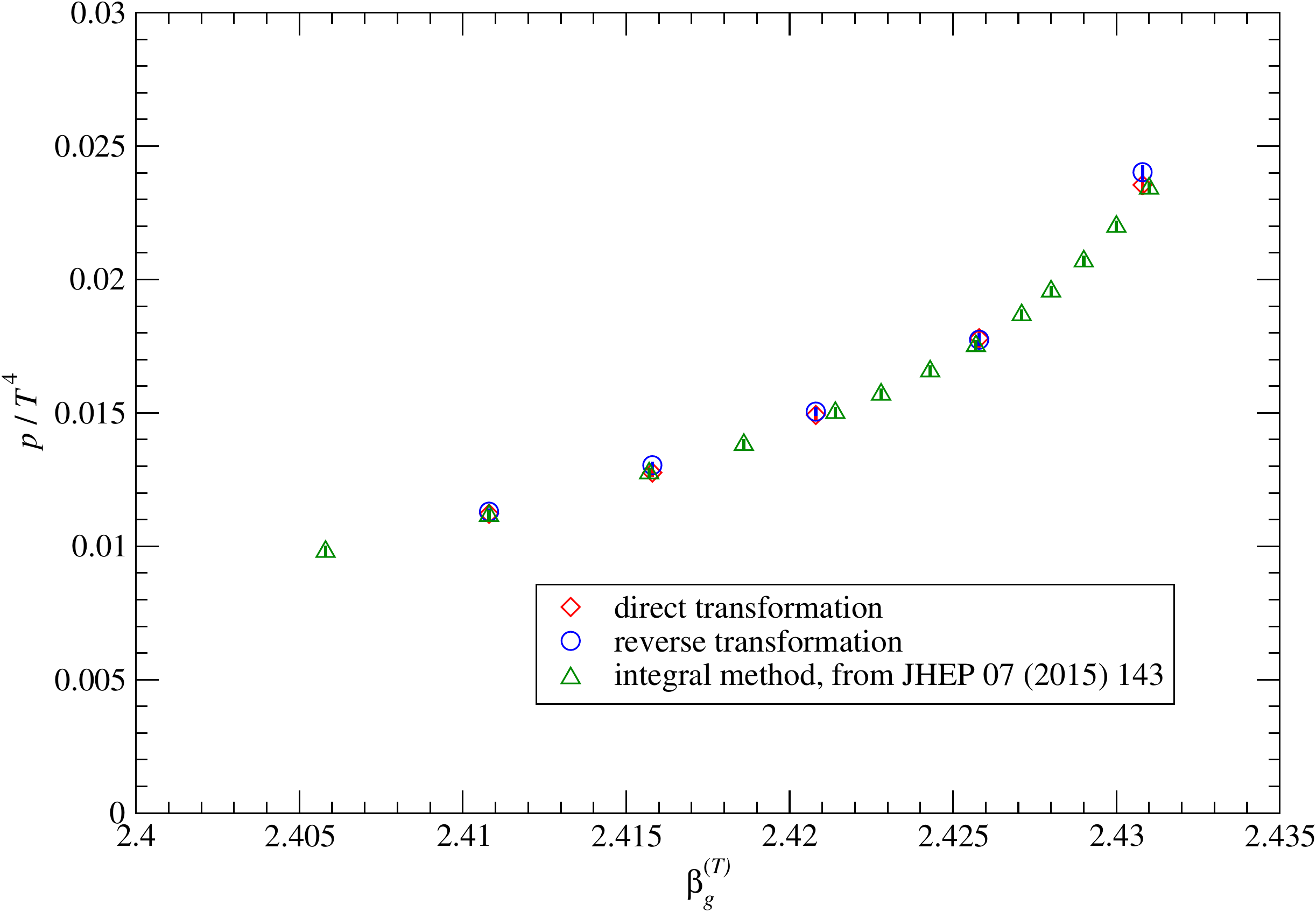}}
\caption{(Color online) Results for the pressure $p$ (in units of the fourth power of the temperature) in the confining phase of $\SU(2)$ Yang--Mills theory, as a function of the Wilson parameter $\betagauge^{(T)}$ (which controls the lattice spacing $a$, and, thus, the temperature $T=1/(aN_0)$), from simulations on lattices with $N_0=6$ and spatial sizes $N_s^3=72^3$ (while the corresponding simulations at $T=0$ were performed on lattices of sizes $\widetilde{N}^4=40^4$). The results obtained using Jarzynski's relation eq.~(\ref{Jarzynski}) with a direct (red squares) and a reverse (blue circles) implementation of the parameter transformation converge to those obtained in ref.~\cite{Caselle:2015tza} using the integral method~\cite{Engels:1990vr} (green triangles).}
\label{fig:su2_pressure}
\end{figure}

Like for the determination of the interface free energy discussed in section~\ref{sec:interface}, we found that, when the transformation is discretized using a sufficiently large number of intervals $N$, the results obtained with a ``direct'' or a ``reverse'' non-equilibrium transformation converge to the same values, which are compatible with those obtained by the integral method used in ref.~\cite{Caselle:2015tza} at nearby temperatures. This is shown in table~\ref{tab:su2_pressure} and in fig.~\ref{fig:su2_pressure}, which report results for the pressure, in units of the fourth power of the temperature, from simulations on lattices at fixed $N_0=6$ (so that the temperature is varied by tuning $\betagauge$, and the results are shown as a function of it) and spatial sizes in units of the lattice spacing $N_s=72$, while the corresponding simulations at $T=0$ were run on lattices of sizes $\widetilde{N}=40$ in all the four directions (so that $\gamma=(6 \times 72^3)/40^4=0.8748$), like in ref.~\cite{Caselle:2015tza}. Note that these results were obtained using independent non-equilibrium transformations from one value of $\betagauge$ to the next, i.e. applying eq.~(\ref{lattice_pressure_Jarzynski}) to compute only the \emph{difference} in pressure. Furthermore, we did not determine the pressure at very small temperatures; instead, we started the analysis at a finite temperature $T_0$, corresponding to $\betagauge=2.4058$, and used the value obtained from the integral method (which is reported in table~\ref{tab:su2_pressure}) for $p(T_0)/T_0^4$ in eq.~(\ref{lattice_pressure_Jarzynski}).

\begin{table}[!htb]
\centering
\begin{tabular}{|c||c|c|c|}
\hline
$\betagauge^{(T)}$ & $p/T^4$, direct & $p/T^4$, reverse & $p/T^4$, integral method \\
\hline
\hline
$2.4058$ &      --       &      --       & $0.00980(22)$ \\
$2.4108$ & $0.01122(9)$  & $0.01130(11)$ & $0.01114(22)$ \\
$2.4157$ &      --       &      --       & $0.01274(22)$ \\
$2.4158$ & $0.01276(15)$ & $0.01304(14)$ &     --        \\
$2.4186$ &      --       &      --       & $0.01381(22)$ \\
$2.4208$ & $0.01492(20)$ & $0.01505(16)$ &     --        \\
$2.4214$ &      --       &      --       & $0.01501(22)$ \\
$2.4228$ &      --       &      --       & $0.01569(22)$ \\
$2.4243$ &      --       &      --       & $0.01656(22)$ \\
$2.4257$ &      --       &      --       & $0.01751(22)$ \\
$2.4258$ & $0.01780(35)$ & $0.01774(24)$ &     --        \\
$2.4271$ &      --       &      --       & $0.01867(22)$ \\
$2.428$  &      --       &      --       & $0.01956(22)$ \\
$2.429$  &      --       &      --       & $0.02068(22)$ \\
$2.43$   &      --       &      --       & $0.02198(22)$ \\
$2.4308$ & $0.02354(37)$ & $0.02402(27)$ &     --        \\
$2.431$  &      --       &      --       & $0.02341(22)$ \\
\hline
$2.4108$ & $0.01122(9)$  & $0.01130(11)$ & $0.01116(51)$ \\
\hline
\end{tabular}
\caption{Results for $p/T^4$ at different values of $\betagauge^{(T)}$ (first column), from simulations on lattices with $N_0=6$ and spatial sizes $N_s^3=72^3$ (while the simulations at $T=0$ were run on lattices of sizes $\widetilde{N}^4=40^4$) using Jarzynski's relation eq.~(\ref{Jarzynski}) with a direct (second column) or a reverse implementation (third column) of the parameter switch, in comparison with those obtained with the integral method~\cite{Engels:1990vr} in ref.~\cite{Caselle:2015tza} (fourth column). The data in the last line provide a comparison of results from the method based on Jarzynski's relation and the integral method, for the same number ($3 \times 10^4$) of gauge configurations. \label{tab:su2_pressure}}
\end{table}

The computational cost to get these results using Jarzynski's relation was rather modest: each of the values of $p/T^4$ reported in table~\ref{tab:su2_pressure} was obtained from simulations with $N=10^3$ (or $N=2 \times 10^3$, at the two largest $\betagauge$ values) and $\nr=30$. Thus, we are in a position to compare the efficiency of the method based on Jarzynski's relation to that of the integral method: for the latter, plaquette expectation values $\langle U_{\Box}\rangle_{\mathcal{T}} $ and $ \langle U_{\Box}\rangle_0$ were calculated using about $10^5$ configurations for each value of $\betagauge$ and then integrated numerically; conversely, using the method based on Jarzynski's relation, each point required either $3 \times 10^4$ or $6 \times 10^4$ configurations, with errors generally comparable to those obtained from the integral method. 

We have to emphasize that a comprehensive comparison in terms of CPU cost between the two methods is not straightforward, since it depends on how many values of $\betagauge$ for which the integrand of eq.~(\ref{lattice_pressure}) is computed are chosen, in order to obtain a reliable numerical integration. To address this issue we attempted a comparison at fixed number of configurations ($3 \times 10^4$) for a single point at $\betagauge=2.4108$: as it can be seen in the last line of table~\ref{tab:su2_pressure}, the statistical uncertainty of the result obtained with the integral method is larger, so that the method based on Jarzynski's relation proves to be computationally more efficient.

Moreover, we remark that, in principle, during a single trajectory $\tin \to \tfin$ it is possible to determine the work (and, hence, the pressure) at any intermediate step between the initial and final value of $\betagauge$, without having to thermalize the system. Even if in this analysis all the values of $p/T^4$ were computed in independent transformations, it is worth stressing that a rather detailed determination of the equation of state would be feasible in this way, provided the correlations among results obtained during a single out-of-equilibrium transformation are properly taken into account.

Finally, we can conclude that the method based upon Jarzynski's relation proved to be very efficient in the determination of the pressure in the temperature region of choice, making it a viable and CPU-cost-effective technique to determine the equation of state.

\section{Discussion and further applications}
\label{sec:conclusions}

In this article, we have shown that the non-equilibrium work relation derived by Jarzynski in statistical mechanics~\cite{Jarzynski:1996ne, Jarzynski:1997ef} can be successfully extended to study problems in lattice gauge theory. This relation links the ratio of the equilibrium partition functions describing a system at two different sets of physical parameters, to the exponential average of the work performed on the system during a non-equilibrium transformation, in which the system parameters and the fields are let evolve.

The generalization of Jarzynski's relation to lattice gauge theory is simply an application of a statistical-mechanics technique to a field-theory context, and does not involve any \emph{ad~hoc} assumptions: this elementary but important point is made clear by the detailed derivation of eqs.~(\ref{generalized_Jarzynski}) and~(\ref{Jarzynski}) in section~\ref{sec:Jarzynski}.

As examples of application, we used Jarzynski's relation to study the interface free energy in the $\Z_2$ gauge model in three dimensions (section~\ref{sec:interface}) and the equation of state in the confining phase of $\SU(2)$ Yang--Mills theory (section~\ref{sec:equation_of_state}).

In the study of the interface free energy, we compared our results with the expectations from effective string theory, and we were able to identify the leading and next-to-leading deviations from the behavior predicted by the Nambu--Got\={o} string. The form of these corrections agrees with theoretical expectations~\cite{Aharony:2013ipa}, but a more detailed quantitative analysis will be carried out later, in a larger-scale study.

In the study of the equation of state, the algorithm successfully reproduced the results obtained with the integral method in ref.~\cite{Caselle:2015tza}, and proved very competitive in terms of computational cost.

In both cases, the calculation of free energies based on this method gave precise results, which converge rapidly to those obtained by different techniques, when the transformation of parameters relating the initial and final partition functions of the system is discretized in a sufficiently smooth way, i.e. when $N$ is large enough: under such conditions, the computational efficiency of the algorithm based on Jarzynski's relation proves to be comparable or, in certain cases, superior with respect to other algorithms.

Numerical calculations involving Jarzynski's relation could also be carried out to study lattice gauge theories coupled to dynamical fermions, including QCD. Although in the present work we have not carried out any studies in this direction yet, there is no conceptual obstruction to generalizing the derivation presented in section~\ref{sec:Jarzynski} to Monte~Carlo calculations involving state-of-the-art fermionic algorithms~\cite{Hasenbusch:2001ne, Luscher:2005rx, Urbach:2005ji, Clark:2006fx, Luscher:2007se, Luscher:2007es}.

In view of the results obtained in the benchmark studies presented here, we envisage a number of further applications of Jarzynski's relation in lattice QCD.

A particularly interesting one could be in studies involving the Schr{\"o}dinger functional~\cite{Symanzik:1981wd, Luscher:1985iu}, which is a powerful method to evaluate running physical quantities in asymptotically free theories~\cite{Luscher:1991wu, Luscher:1992an}. The Schr{\"o}dinger functional provides an elegant, gauge-invariant, finite-volume scheme, which is free from many of the technical challenges related to the chiral limit or to the presence of bosonic-field zero-modes for theories defined on a torus, and which, in addition, is particularly suitable for perturbative computations. 

In this approach, one considers the evolution of the system during a Euclidean-time interval $L$, from an initial state $\mathcal{I}$, to a final state $\mathcal{F}$. At the classical level, the action $\Scl$ of the field configuration induced by the presence of these boundary conditions is inversely proportional to the squared bare coupling of the theory $g$. At the quantum level, denoting the Hamiltonian of the system by $H$, one can compute the transition amplitude
\begin{equation}
Z_{\mathcal{I},\mathcal{F}; L} = \langle \mathcal{F} | \exp \left(- H L \right) | \mathcal{I} \rangle
\end{equation}
and define the effective action $\effaction$ as
\begin{equation}
\effaction = - \ln Z_{\mathcal{I},\mathcal{F}; L}.
\end{equation}
Then, one can define a renormalized coupling $\bar{g}$ at the momentum scale $L^{-1}$, by assuming that $\effaction$ is proportional to $1/\bar{g}^2$. In practical simulations, if the boundary states $\mathcal{I}$ and $\mathcal{F}$ depend on a set of parameters $\chi$, then $\bar{g}^2(L^{-1})$ can be computed from
\begin{equation}
\label{SF_coupling}
\bar{g}^2(L^{-1}) = g^2 \frac{\Scl^\prime}{\effaction^\prime},
\end{equation}
where the prime denotes derivation with respect to $\chi$.

This approach has been used for studies of pure-glue non-Abelian gauge theories~\cite{Luscher:1992zx, Luscher:1993gh, Bode:1998hd, Lucini:2008vi} and can be extended to include dynamical fermions~\cite{Sint:1993un}: this has direct applications in QCD~\cite{Sint:1998iq, Bode:1999sm} and in strongly interacting theories~\cite{Appelquist:2007hu, Shamir:2008pb, Hietanen:2009az, Karavirta:2011zg, DeGrand:2011qd, DeGrand:2012qa} that might provide viable models for dynamical breaking of electro-weak symmetry at the TeV scale~\cite{Sannino:2009za, DelDebbio:2010zz} and/or for composite dark matter~\cite{Kribs:2016cew}.

Using Jarzynski's relation, in principle one could evaluate $\bar{g}^2(L^{-1})$ by computing the variation induced in $Z_{\mathcal{I},\mathcal{F}; L}$ by a change in the $\chi$ parameters that specify $\mathcal{I}$ and $\mathcal{F}$.

Finally, it is tempting to think that Jarzynski's relation could also find applications in lattice QCD at finite density, where, as we mentioned in section~\ref{sec:introduction}, the loss of $\gamma_5$-Hermiticity of the Dirac operator induces a severe sign problem~\cite{deForcrand:2010ys, Philipsen:2012nu, Levkova:2012jd, Aarts:2013lcm, D'Elia:2015rwa, Gattringer:2016kco}. In particular, the connections between Jarzynski's relation and the reweighting technique~\cite{Ferrenberg:1988yz, Barbour:1997bh, Fodor:2001au}, that we mentioned in section~\ref{sec:Jarzynski}, deserve further investigation. We leave these issues for future work.

\vskip1.0cm 
\noindent{\bf Acknowledgements.}\\
The simulations were run on INFN Pisa GRID Data Center and on CINECA machines. The work of A.~T. is partially supported by the Danish National Research Foundation grant DNRF90. We thank C.~Bonati, R.~C.~Brower, M.~D'Elia, M.~Mesiti, M.~Pepe, A.~Ramos and E.~Vicari for helpful comments and discussions.

\bibliography{paper}

\end{document}